\documentclass[12pt,preprint]{elsarticle}
\usepackage{fullpage}
\usepackage{graphicx,amsmath}
\newcommand{\be}{\begin{equation}}
\newcommand{\ee}{\end{equation}}
\newcommand{\beqa}{\begin{eqnarray}}
\newcommand{\eeqa}{\end{eqnarray}}
\newcommand{\hf}{{1\over 2}}
\newcommand{\du}{\partial _u}

\newcommand{\rg}{\sqrt{-g}}
\newcommand{\lb}{\left(}
\newcommand{\rb}{\right)}
\numberwithin{equation}{section}
\def\f{\frac}
\def\pa{\partial}
\def\non{\nonumber}

\def\d{\delta}

\def\l{\lambda}

\def\o{\omega}

\def\r{\rho}
\def\s{\sigma}
\def\t{\tau}

\graphicspath{{figuresbcs/}}
\begin{document}
\begin{frontmatter}
\title{Strong Coupling BCS Superconductivity and Holography}
\author[imsc]{S. Kalyana Rama}
\ead{krama@imsc.res.in}
\author[du]{Swarnendu Sarkar} 
\ead{ssarkar@physics.du.ac.in }
\author[imsc]{B. Sathiapalan} 
\ead{bala@imsc.res.in}
\author[tifr]{Nilanjan Sircar\corref{cor1}}
\ead{nilanjan@theory.tifr.res.in}
\address[imsc]{The Institute of Mathematical Sciences, Taramani, Chennai, India 600113}
\address[du]{ Department of Physics and Astrophysics, University of Delhi, Delhi 110007, India}
\address[tifr]{Tata Institue of Fundamental Research, Colaba, Mumbai 400 005, India }
\cortext[cor1]{Corresponding author}
\fntext[fn1]{IMSC/2011/04/2 , TIFR/TH/11-14,arXiv:1104.2843 [hep-th] }    
\begin{abstract} 
We attempt to give a holographic description of the microscopic
theory of a BCS superconductor. Exploiting the analogy with
chiral symmetry breaking in QCD we use the Sakai-Sugimoto model
of two D8 branes in a D4 brane background with finite baryon
number. In this case there is a new tachyonic instability which
is plausibly the bulk analog of the Cooper pairing
instability. We analyze the Yang-Mills approximation to the
non-Abelian Dirac-Born-Infeld action. We give some exact solutions of
the non-linear Yang-Mills equations in flat space and also give
a stability analysis, showing that the instability disappears in
the presence of an electric field. The holographic picture also
suggests a dependence of $T_c$ on the number density which is
different from the usual (weak coupling) BCS. The flat space
solutions are then generalized to curved space numerically and
also, in an approximate way, analytically. This configuration
should then correspond to the ground state of the boundary
superconducting (superfluid) ground state. We also give some
preliminary results on Green functions computations in the Sakai
- Sugimoto model without any chemical potential.
\end{abstract}
\end{frontmatter}
\section{Introduction}
    
The application of AdS/CFT techniques to strongly correlated
systems in condensed matter situations is a very promising
development \cite{HKSS, DHS, G, GPR, HHH, LMV, FLMV,HLM,KZ}. This is
because there are genuinely strong coupling fixed points in
condensed matter systems in contrast to particle physics. Even
the so called strong interactions of particle physics are
described by an asymptotically free theory, QCD, which has a
fixed point at zero coupling. Of course the coupling constant of
QCD is large at low energies and it may be that to a large
extent these strong coupling regions dominate the dynamics in
certain phenomena. In that case it is plausible that one can
borrow results from N=4 super Yang-Mills also at strong
coupling, where indeed one can apply AdS/CFT. Presumably the
agreement between the small values of $\frac{\eta}{s}$ seen in
quark gluon plasma in heavy ion collisions and in N=4 Yang-Mills
has to be understood on these lines~\cite{KSS,CLMR}.
 
In applying AdS/CFT techniques to condensed matter physics or
QCD one can take a phenomenological approach where one appeals
very strongly to universality. This means that the details of
the model in the UV are unimportant for low energy
phenomena. Many recent papers have analyzed this idea and
provided very useful insights in understanding earlier
calculations \cite{FP, NS, FLR, HP}. This may be qualitatively
true in many situations. Thus there are some gravity
descriptions of QCD-like theories where features such as
confinement or chiral symmetry breaking may be seen. They can be
said to be in the same universality class as QCD. They are not
known to be reliable quantitatively (and not in the UV regime
either). Thus the glueball spectrum is not in very good
agreement with lattice results although there are some generic
features that are the same\cite{COOJ, BMT}. But it may even be
possible to do this in a quantitative way by introducing some
parameters and fitting to experimental data.  Thus in standard
field theory QCD - chiral Lagrangians are motivated by this
idea. This is also the spirit of the Landau-Ginzburg analysis in
critical phenomena in condensed matter physics.  In the context
of AdS/CFT there have been some approaches to QCD with this
philosophy\cite{Brodsky}.
 
 However in condensed matter systems there are models that one
takes a little more seriously. There are models such as the
Ising model or Hubbard model, that one would like to solve as a
fully quantum theory.  This means that while the model may be an
approximation to the condensed matter physics, the model itself
as a mathematical system is taken fully seriously. One cannot do
this for instance in Landau-Ginzburg theory or chiral
Lagrangians - both of which are non renormalizable. Perturbative
non renormalizability implies that order by order one has to
extend the model by adding a large number of additional
interactions and consequent additional free parameters. The
model then has to be understood as an effective low energy
theory, with a cutoff. Only then does it make sense as a quantum
theory.
 
This sort of distinction applies to AdS/CFT models as well. The
models that start with a bulk gravity theory and where one does
classical gravity calculations to extract physics of the
boundary belong to the Landau-Ginzburg/chiral Lagrangian
class. One cannot take them seriously beyond the large N limit,
because the quantum gravity theory is not well defined. Thus one
does not take them seriously as a mathematically well defined
system beyond leading order. Furthermore the boundary theory has
no independent definition. Very often we do not know the action
or even the field content.
 
On the other hand there are models where one begins with a
string theory background, where in principle the theory is
defined at the quantum level.  Some examples are the models for
QCD first studied by Witten \cite{EW1} and the more recent
versions of it such as the Sakai-Sugimoto model, where flavor
D-branes are added and give fundamental fermions - quarks
\cite{SS1, KK, KMMW, SS, EKSS, KKSS, AHJK, DN, ASY, BSS}.
Although these theories are non conformal, there is an
underlying $AdS_7$ geometry and hence one can think of these
theories as deformations of some conformal field theory. Apart
from being well defined quantum mechanically, they have the
advantage that the field content of the boundary theory is
precisely known. One of the nice features of the Sakai-Sugimoto
model is that it gives a geometric picture of chiral symmetry
breaking.  Superconductivity in strongly
coupled $\mathcal{N}=2$ Super Yang-Mills theory with
fundamental matter was studied via AdS/CFT in~\cite{AEKK}

It is well known that the instability of Cooper pairing in
superconductors is very similar to that of chiral symmetry
breaking \cite{Nambu}\footnote{ This is obvious from the title
of the paper of Y. Nambu and G. Jona-Lasinio: \\ ``Dynamical
Model of Elementary Particles Based on an Analogy with
Superconductivity". }. This analogy motivates us to study the
Sakai-Sugimoto model to understand strong coupling BCS
superconductivity. The Sakai Sugimoto model has the feature that
supersymmetry is broken and the lightest modes in the boundary
are fermions in the fundamental representation, rather than
scalars or some supermultiplet of particles (as in some of the
D3-D7 models). The fact that the light particles are fermions,
makes this realistic as one can be assured that what is
condensing is a bound state of fermions rather than some
scalars. Thus one is actually seeing the Cooper pairing. This
goes beyond the Landau Ginzburg description of the theory as a
$U(1)$-Higgs system.
 
{\bf BCS vs Chiral Symmetry Breaking:}
 
There are two important differences between the Cooper pairing
instability in BCS theory and chiral symmetry breaking. Both
these must be kept in mind while using the Sakai-Sugimoto
model. One is the presence of a Fermi surface in the BCS
situation. Chiral symmetry breaking as studied by
Nambu-JonaLasinio (NJL) takes place in the vacuum. This
difference shows up in the two gap equations. The NJL gap
equation is a one loop tadpole vanishing condition. The one loop
fermion contribution is balanced against a tree level constant
and gives:
\be  \label{NJL}
g \int_0^\Lambda d^3k ~ {1 \over \sqrt {k^2 + m^2} } \approx 1
\ee
where $m >0$ is the sign of chiral symmetry breaking. The
analogous BCS gap equation is
\be  \label{BCS}
g \int_{k_F}^{k_F + \delta} d^3k ~ 
{1 \over \sqrt {\epsilon _k^2+ \Delta ^2}} 
\approx 1
\ee
where $\epsilon _k = {(k^2 - k_F^2) \over 2 m}$

If we put $m = 0$ in (\ref{NJL}) and solves for $g \;$, one
finds $g \equiv g_{cr} \;$. For $g > g_{cr}$ one finds $m > 0
\;$. This is the chiral symmetry breaking phase. In the original
NJL calculation $g_{cr} \approx {2 \pi^2 \over \Lambda^2}
\;$. Contrast this with (\ref{BCS}). Only states above the Fermi
sea contribute and the criterion for the instability is $\Delta
> 0 \;$. But note that $\Delta$ is measured from the Fermi
surface. If we set $\Delta = 0$ there is an IR divergence as $k
\rightarrow k_F \;$, which in turn is because the measure $d^3 k
\approx 4 \pi k_F^2 dk$ does not vanish. Thus for {\em
arbitrarily small} $g$ one can find a solution with $\Delta > 0
\;$. In fact one can write (\ref{BCS}) as
\[
g \; {d n \over d \epsilon} \; \int_0^{\epsilon_c} 
~ d \epsilon {1 \over \sqrt{\epsilon^2 + \Delta^2}} \; \; . 
\]
For $g \; {d n \over d \epsilon} << 1 \;$, the solution is
$\Delta \approx \epsilon_c e^{ - {1 \over g \; {d n \over d
\epsilon}}} \;$. On the other hand for $g \; {d n \over d
\epsilon} >> 1 \;$, we have $\Delta \approx \epsilon_c g \; {d n
\over d \epsilon} \;$. The dependence of the gap $\Delta$ and
hence also the critical temperature on the parameters $(g , \;
{d n \over d \epsilon} )$ is very different. In weak coupling
there is a non perturbative dependence. In strong coupling
(which is what the holographic calculation gives) the
temperature is proportional to some power of the number density.

Thus in the presence of a Fermi surface there is always a BCS
instability.\footnote{This can be also be seen using the
Renormalization Group formalism where the BCS perturbation shows
up as a relevant operator \cite{S}.}  This points to an
important modification of the Sakai-Sugimoto model. We need to
include a chemical potential for ``baryon" number {\em i.e.} we
need a finite number density of baryons in the boundary theory.
 
The second important difference is that it is a chiral rather
than vector symmetry that is being broken in the NJL
model. Chiral symmetry is broken as soon as the fermion has a
bare mass, so in this sense it is a symmetry that is easily
broken explicitly. In fact the Vafa-Witten theorem shows that
vector symmetries are not broken spontaneously in QCD-like
theories when the bare quark masses are non zero\cite{VW}.
 
In QCD and also the Sakai-Sugimoto model the vector symmetry is
unbroken. {\em With zero bare mass} for the quarks, the unbroken
phase (with two flavor D8-brane pairs) has $SU(2)_L \times
SU(2)_R = SO(4) \;$. The quark bilinears $\bar q^i q_j$ form a
four-vector under this $SO(4)$. When one component gets an
expectation value what is left unbroken is {\em by definition}
the vector $SU(2)$ - also called strong isospin. A basis can be
chosen such that the Goldstone bosons are $\bar q \vec \sigma q$
- the pions. In this situation the question of whether this
condensate could have broken the $SU(2)$ vector is meaningless.
In fact if we find that one of the components of the pions
condense due to quantum effects, this corresponds to a
realignment of the vacuum and one can perform an $SO(4)$
rotation so that what is left unbroken is still an $SU(2)
\;$. If there is a bare quark mass then there is a well defined
notion of which is the vector symmetry. This is the situation
studied by Vafa-Witten. Also the presence of electromagnetism
changes the situation, because there is a preferred axis defined
by the charge of the pions and this breaks the strong isospin.
Thus if due to external field effects the $\langle \pi^+ \rangle
\neq 0$ then we cannot rotate this away and we have a breaking
of $U(1)_{em}$. \footnote{Note that if $\langle \pi^0 \rangle
\neq 0 \;$, when the quarks are massless, this can be rotated
away. Thus if $u_L^\dagger u_R + d_L^\dagger d_R = v_1$ (the
isosinglet) and $u_L^\dagger u_R - d_L^\dagger d_R = v_2$ (the
neutral component of the isotriplet), then an $SU(2)_L$ rotation
that commutes with $U(1)_{em}$ can rotate one into the other.}

Thus if we are to spontaneously break a vector symmetry then the
Vafa-Witten theorem tells us that it cannot be in the vacuum.
We must have a finite number density of fermions. This again
points to the same requirement. The Sakai Sugimoto model has to
be modified: we need a chemical potential and a finite number
density. Such modification have been studied, see \cite{BLL,
CDKLY} for this model and \cite{KMMMT} for other models.

Baryon number corresponds to a $U(1)$ charge corresponding to a
gauge field on the $D8$ brane. To probe this we need a field on
the $D8$ brane that has non-zero baryon number. We can achieve
this if we have two $D8$ branes, so that the $U(1)$ can be
embedded in $SU(2) \;$. The charged gauge fields (``$w^\pm$") or
charged scalar can then be used as the fields dual to the baryon
number violating charged condensate. If the branes are separated
this would correspond to a Georgi Glashow model where the
$SU(2)$ is broken to a $U(1)_3$. (The scalar field on the D8
brane, corresponding to its transverse fluctuations, becomes the
adjoint scalar field of the Georgi Glashow model.) The generator
of this $U(1)$ will be called $t_3$ defined later in this
section. This makes classical computations reliable and so we
will implement this.

By an AdS/CFT--type dictionary one would expect that the
boundary value of the charged field would correspond to the
value of the condensate. \footnote{The asymptotic geometry here
is not AdS. However the underlying M-theory geometry has an
$AdS_7$ and so while the usual AdS/CFT dictionary does not
apply, one may expect some modified version of this to hold. A
similar situation involving D1 branes and an underlying $AdS_4$
has been studied in \cite{DMTW}.}

In the boundary theory one has two quarks in the fundamental of
$N_c$ and a doublet of flavor, which we denote by $u,d \;$. If
we turn on a chemical potential corresponding to the 3-component
of $SU(2)$ this couples to $u_L^\dagger u_L - d_L^\dagger d_L
\;$.  The height of the Fermi surface will then be different for
$u$ and $d$.  But in the confined phase one should define Fermi
surfaces for color singlet fermions. The natural candidates are
the baryons (for $N_c$ odd). Thus one expects an excess of
baryons with $t_3 > 0 \;$\footnote{$t_3$ is defined in the next page}. (If $N_c$ is even this has to be
interpreted as a chemical potential for scalar baryons.  This is
more like a BEC situation.) Then there is the question of what
the BCS condensate consists of.  Since in the bulk we are
studying only the fields on the left branes, we are studying
boundary fields involving $u_L ,d_L \;$. \footnote{This assumes
that one can treat the dynamics of the left and right brane
independently. This is not the situation in the Sakai-Sugimoto
model which is left-right symmetric. However in more general
backgrounds with electric fields this is possible.}  One can
expect pairing between the $u_L,d_L$'s of the form $u_Lu_L$ or
$d_L d_L$. This would break color and give rise to color
superconductivity \cite{ARW}. Since our bulk probes do not carry color we
cannot study this easily. Thus we focus on color singlets --
mesons and baryons. One cannot make scalar mesons out of
$u_L,d_L$ alone - the scalar mesons necessarily involve both
$u_L$ and $u_R \;$.\footnote{The vector mesons cannot condense
without breaking Lorentz invariance. One could imagine a
bilinear product of vector mesons condensing.} Note that $\bar u
d = u_L^\dagger d_R + d_L^\dagger u_R \;$. The massless
pseudo-scalar mesons, the pions are frozen by our boundary
conditions on the gauge fields. The other mesons are massive. In
principle they can still condense, but being made out of $u_L$
and $u_R$ cannot be probed by our bulk fields which are only on
the left brane. On the other hand baryons can form Cooper pairs
and condense as in BCS theory. Thus if $N_c$ is odd $B = \epsilon_{a_1, a_2,
\cdots, a_{N_c}} u_L^{a_1} d_L^{a_2} u_L^{a_3} d_L^{a_4} \cdots
u^{a_{N_c}}$ and one can have $\langle BB \rangle \neq0$. 
\footnote{If $N_c$ is even, the Baryons are scalars and can
condense directly. This is like a BEC.}  Even though the gauge
theory is strongly coupled, the coupling between the color
singlet baryons is not so obviously strong. Thus one need not
expect protons and neutrons to form Cooper pairs and condense in
the vacuum. It certainly doesn't happen in real life
QCD. However with a finite chemical potential and a Fermi
surface for baryons it is possible. This description in terms of Baryons
is reminiscent of the complementary description of color superconductivity
in \cite{ARW}. This is the most likely
boundary dual of the bulk condensate that we are studying.

\begin{figure}[htbp]
\begin{center}
\includegraphics[scale=.6]{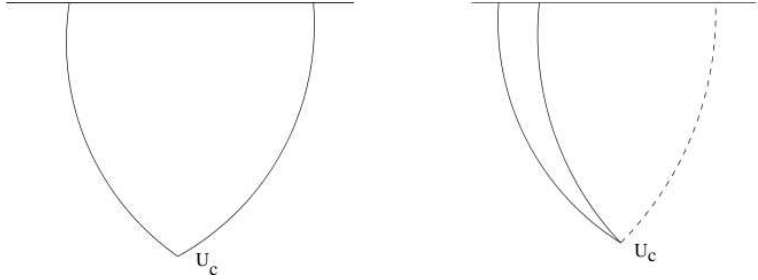}
\caption{In the presence of a point source of baryon number
there is a cusp singularity as shown on the left. When we have
two branes and $SU(2)$ breaking source charge the branes
separate. In the second figure the mirror D8 branes are not shown in detail but are indicated by a dotted line. Their configuration is chosen to satisfy the force balance condition. }
\label{cusp}
\end{center}
\end{figure}

From the bulk perspective adding a point source baryon charge,
corresponding to the overall U(1), to the Sakai-Sugimoto model
makes the D8 branes have a cusp geometry \cite{BLL} (see Figure
(\ref{cusp})).

The coupling of this point charge to the $U(1)$ gauge field on
the brane is via a Chern-Simons term in the action via a point
like instanton configuration.  The instanton can also be
understood as a D4 brane wrapped around the $S^4$ and has the
same action \cite{SS1}. However the instanton coupling requires
more than one brane since it is a configuration involving non
Abelian gauge fields.  Thus we should interpret our single brane
as a collection of $N_d$ coincident branes, with $N_d\ge 2$. The only role of the
extra branes is to accommodate the instanton configuration.
Otherwise as far as the rest of the dynamics that we are
interested in is concerned, the collection behaves as one $U(1)$
brane. This $U(1)$ is the overall center of mass $U(1)$ in the
$U(N_d)$.  All this is exactly as in \cite{BLL}.

In our case the new thing is that we have two such D8 branes
with a $U(1)$ charge and Chern Simons term for each brane. Thus
we have a $U(2)$ gauge symmetry on the branes. When the charges
on the two branes are unequal then the branes come in at
different angles and separate. The $U(2)$ symmetry is broken
(Higgsed) to $U(1)\times U(1)$ when the branes are
separated. These can be denoted as $U(1)_B\times
U(1)_3$. $U(1)_B$ is the overall $U(1)$ of the $U(2)$ and
$U(1)_3$ corresponds to the generator $t_3$ of the
$SU(2)$. 

We define the $SU(2)$ generators $(t_+, t_-, t_3) \;$:
\be
t_+ = \left( 
\begin{array}{cc}
0 & 1 \\
0 & 0 
\end{array} \right) 
~~~~~~ t_- = \left( 
\begin{array}{cc}
0 & 0 \\
1 & 0 
\end{array} \right)
~~~~~~ t_3 = \frac{1}{2} \; \left( 
\begin{array}{cc}
1 & 0 \\
0 & - 1 
\end{array} \right)
\ee
which satisfy the $SU(2)$ algebra 
\[
[t_3, t_{\pm}] = \pm \; t_{\pm} , ~~~~
[t_+, t_-] = 2 \; t_3 ~~~~
\{t_+, t_-\} = I \; \;. 
\]

 As mentioned
above the Chern Simons coupling of each brane comes from having
$N_d$ coincident branes. Thus we actually have $2N_d$ branes and
a $U(2N_d)$ symmetry.  The $U(2)$ group that we are concerned
with in this paper is thus embedded in an obvious way in the
$U(2N_d)$ group. (The mathematical description of this embedding
is given for completeness in  Appendix B.) The brane
configuration thus leaves unbroken a $U(1)_B \times SU(N_d) \times
U(1)_3 \times SU(N_d)$ group.
 
From the bulk brane dynamics also we expect a condensate,
because it is known that when branes intersect at an angle, in
general there are tachyonic excitations \cite{BDL, HT, N}. Thus
there should be a condensate of charged fields. {\em This should
then correspond to the Cooper pairing instability of the
boundary theory.} \footnote{We are assuming that there are no
other instabilities in the boundary.} This generates a non zero
electric field. In the presence of a non-zero electric field and
charged condensate the tachyonic mode should get lifted and one
expects a stable solution with electric field. This condensate
breaks one more $U(1)$, namely $U(1)_3 \;$, in the $U(1)_B
\times SU(N_d)\times U(1)_3 \times SU(N_d)$ group left unbroken
by the geometric description of the D8 branes. As in \cite{BLL}
since the $SU(N_d)\times SU(N_d)$ gauge fields play no role in
any of the dynamics we do not mention them again in the
discussions below.

\begin{figure}[htbp]
\begin{center}
\includegraphics[scale=.5]{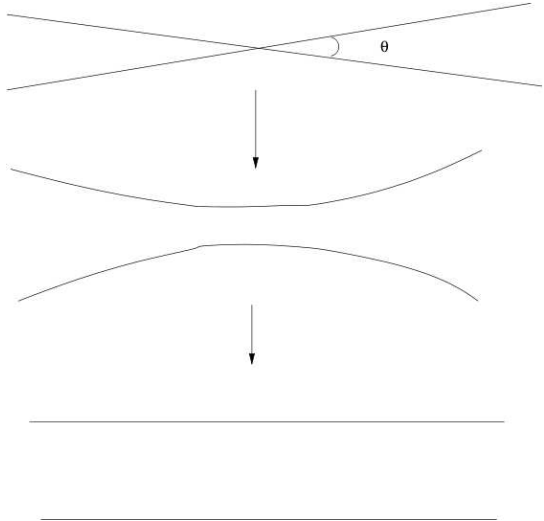}
\caption{Tachyonic instability}
\label{tach}
\end{center}
\end{figure}

\begin{figure}[htbp]
\begin{center}
\includegraphics[scale=.5]{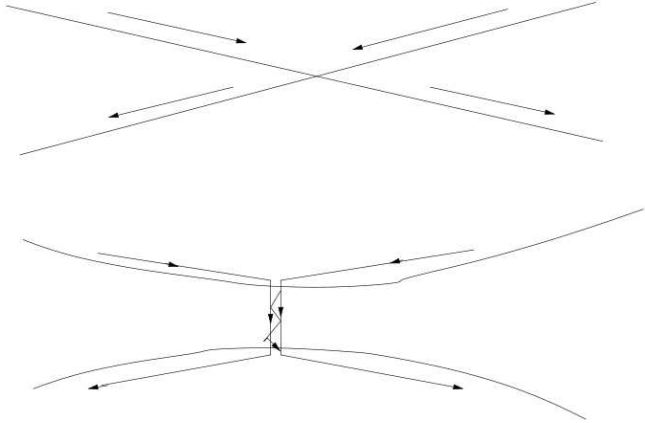}
\caption{Electric field causes open strings to be formed
that stabilize the configuration.}
\label{E}
\end{center}
\end{figure}

There is a geometric interpretation of the instability that
makes this clear (see Figures \ref{tach}, \ref{E}). The
tachyonic mode corresponds to the tendency of the D branes to
separate in the region around the intersection, and become
parallel, because this reduces the energy. However, once there
are open strings connecting the branes, it costs energy to
separate the branes and make them parallel. Thus one expects
stable solutions with condensates. We also expect to find
analytic solutions in flat space describing charged
condensates. The absence of tachyons in the presence of an
electric field is confirmed by a perturbative analysis of the
fluctuation spectra.

In fact there are solutions to the Yang-Mills equations that
describe precisely such a situation: as soon as we have a non
zero $t_3$ charge, there are solutions where $w^\pm$ fields
condense.  If the condensate extends to the boundary, we expect
superconductivity in the boundary. We also give a proposal for
calculating Greens functions in this situation. The complication
here is that the branes do not extend all the way to the black
hole. Hence the usual ingoing boundary conditions at the horizon
cannot be applied because there is no horizon crossing. We apply
this technique to the Sakai-Sugimoto model where Greens
functions have not been calculated. We expect that the boundary
fermions being massive the Green's function (say, of the
currents) should show a gap in the imaginary part. We find that
the conductivity calculated using this prescription seems to
have the expected properties. However we leave a detailed
analysis to the future. Also applying to the BCS situation is
more complicated and we have not attempted it in this paper.

We can summarize the above discussion as follows: In the bulk we
have two (sets of coincident) D8 branes in the configuration
shown in Figure 1. There are two $U(1)$ charges $d_1,d_2$
corresponding to each brane. This can be represented as
\be  \label{u1}
\left( \begin{array}{cc} d_1 & 0 \\ 0 & d_2 \end{array} \right)
= d_0~ {\mathbf I} + d_3 ~t_3 = 
d_0 \left( \begin{array}{cc} 1 & 0 \\ 0 & 1 \end{array} \right)
+ {d_3 \over 2} \left(
\begin{array}{cc} 1 & 0 \\ 0 & -1 \end{array} \right) \; \; \; . 
\ee
$d_0$ is the $U(1)_B$ baryon number charge and $d_3$ is the
$U(1)_3$ (corresponding to $t_3$) charge.  When $d_3>0$ the
branes come in at different angles at the cusp $u_c$. For
$u>>u_c$ the branes are parallel but separated by a finite
distance. The $SU(2)$ is then broken to $U(1)_3$ as in the
Georgi-Glashow model as soon as $d_3>0$.  As long as no off
diagonal $SU(2)$ fields are turned on the DBI action is exactly
that of two independent branes and is easily analyzed. The free
energy of this configuration can be studied for arbitrary values
of $d_0,d_3$.

We have also seen above that in this situation, a flat space
analysis shows that there is a tachyonic instability which we
plausibly identify with the BCS instability - assuming there are
no other new unexpected boundary instabilities.

To quantitatively analyze this condensate in the D4 brane
background we restrict ourselves to $d_3<< d_0$ and all $SU(2)$
fields smaller than the other terms in the DBI action. In this
case it is consistent to expand the DBI action and keep the
Yang-Mills part. A flat space analysis of Yang-Mills also tells
us that there is a stable ground state where some charged fields
condense. This encourages us to explore solutions of the
Yang-Mills equations of motion in the D4 brane background. The
curved space equations are then solved exactly, numerically and
in some approximation analytically. There are solutions where
the fields are small, validating the truncation of the DBI
action to the leading Yang-Mills part. Also the existence of a
charged condensate breaking the $U(1)_3$ symmetry implies that
this approximation captures qualitatively the phenomenon that we
are trying to describe. The smallness of the fields assures us
that higher order terms can only change things quantitatively by
small amounts - not qualitatively. Thus we have a self
consistent scheme that gives us a ground state with a charged
condensate. The $U(1)_3$ symmetry is broken. Thus we expect the
same symmetry breaking in the boundary also. This is then
identified with the BCS phenomenon.

One also expects that at finite temperature the positive
$(mass)^2$ due to temperature effects will overwhelm the
tachyonic tendency of the intersecting brane. The negative
$(mass)^2$ is proportional to the angle between the branes,
which in turn is fixed by the number density. So at a
(dimensionless) $T_c$ roughly equal to the number density we
expect superconductivity to disappear. This dependence on the
power of number density is reminiscent of the strong coupling
expression given earlier (for the gap). However we need a
detailed calculation before anything concrete can be said. In
any case it is very different from the weak coupling BCS
expression.

This paper is organized as follows: In Section 2 we describe the
brane solution in the presence of charges. In Section 3 we
describe the Yang-Mills solution in flat space with $w^\pm$
condensate relevant for the D4-D8 case and analyze its stability
In Section 4 we give the solutions (exact numerical and
approximate analytic) again in the D4 background metric. This
establishes the existence of a charged condensate.  In Section 5
we study analytically the phase structure at zero temperature,
i.e. as a function of chemical potential but without any
condensate. With a condensate we give some numerical
results.(The finite temperature analysis will be done
elsewhere.) In Section 6 we give a sample calculation of a
Greens function in the Sakai-Sugimoto model using a new
prescription. We conclude in Section 7 with a discussion of our
results and some open questions. The Appendices contain a
collection of some useful results that are helpful in obtaining
the equations.

\section{Background Brane Configuration}

The ten dimensional $D4$ brane background metric is
\[
ds^2 = \left( {U \over R} \right)^{3 \over 2} 
( \eta _{\alpha \beta} dy^\alpha dy^\beta + f d\theta^2)
+ \left( {R \over U} \right)^{3 \over 2} \;
({dU^2 \over f} + U^2 d\Omega_4^2)
\]
\be 
e^\phi = g_s \left( {U \over R} \right)^{3 \over 4} , ~~~~
dC_3 = {2 \pi N_c \over \Omega_4} \epsilon_4 , ~~~~
f = 1 - {U_{KK}^3 \over U^3} 
\ee
Here $y^\alpha (\alpha = 0,1,2,3)$ and $\theta$ are the D4-brane
directions. $d\Omega_4^2$ is the line element of $S^4$,
$\epsilon_4$ is the volume form in 4 dimensions, and $\Omega_4=
{8 \pi^2 \over 3}$ is the volume of a unit 4-sphere.  $C_3$ is
the dual gauge field of the D4-brane and has components along
$S^4 \;$. $R^3 = \pi g_s N_c l_s^3$ sets the scale of the space
time curvature. $U_{KK}$ is related to the supersymmetry
breaking scale, which is the period of the $\theta$ coordinate,
${4 \pi \over 3} R^{3 \over 2} \; U_{KK}^{- {1 \over 2}} \equiv
2 \pi R r_4 \;$.

One can define new dimensionless coordinates: $u = {U \over R}$,
$x^\alpha = {y^\alpha \over R}$ and $\tau = {\theta\over R}$, as
well as a dimensionless gauge field: $a_\mu = {2 \pi \alpha'
\over R} A_\mu \;$. In these units, the period of the $\tau$
coordinate is ${4 \pi \over 3} R^{1 \over 2} \; U_{KK}^{- {1
\over 2}} \equiv 2 \pi r_4 \;$, the metric becomes
\[
ds^2 = R^2 \; [ u^{3 \over 2} 
(\eta _{\alpha \beta} dx^\alpha dx^\beta
+ f d\tau^2) + u^{- {3 \over 2}} \; ({du^2 \over f}
+ u^2 d\Omega_4^2)]
\]
where $f(u) = 1 - {u_{KK}^3 \over u^3}$ with $u_{KK}={U_{KK}
\over R} \;$, and $e^\phi = g_s \; u^{3 \over 4} \;$.

Our configuration consists of two flavor D8 branes with a
background gauge field and some delta function sources. (As
explained in the Introduction, these two D8 branes are actually
two sets of $N_d$ number of coincident D8 branes.) The delta
function sources correspond to baryons that are uniformly
distributed in the $(x^1, x^2, x^3)$ directions and at fixed $u
= u_c \;$. Baryon number can come from D4 branes wrapped around
$S^4$ and immersed in the D8 brane. They can equally well be
thought of as an instanton background gauge field configuration
of the non Abelian gauge field on the D8 brane. As shown by
Sakai and Sugimoto, either picture yields the same value for the
action.

The Chern-Simons action is (we have integrated by parts the
action $C_3\wedge F\wedge F\wedge F$ and separated a $U(1)$ part
for $A_0$)
\[
{1\over 16 \pi^2} \int _{S^4} dC_3 
\int _{R^4} F\wedge F  \int_{S^1} d t_E A_0 
\]
with $\int_{S^4} dC_3 = 2 \pi N_c$ and $\int_{R^4} F \wedge F =
8 \pi^2 N_4 \int du \delta(u-u_c)$ where we have assumed the
instantons are localized at $u_c$.

As mentioned in the introduction even when we have one D8 brane we have to think of it as a set of coincident D8 branes 
with the associated non Abelian gauge fields providing the localized instanton configuration~\cite{BLL,SS1}. 
These non Abelian fields are zero everywhere else and play no role in the rest of the dynamics. Thus we assume
following Appendix B that $A= A^a \lambda ^a + A \lambda ^0$, where $\l ^a$ are the Gell-Mann $\l$-matrices for $SU(3)$ and $\l^0$ is the $3\times 3$ identity matrix. The $SU(3)$ part of the gauge field is assumed to correspond to a localized instanton, so that $\int_{R^4} F \wedge F =
8 \pi^2 N_4 \int du \delta(u-u_c)$. $A_0= A_0 \l ^0$ above is then assumed to be the $U(1)$ part.

When we have two $D8$ branes we assume two decoupled sets of the above, each with a $U(3)$ gauge group. The embedding of this $U(3)\times U(3)$ in $U(6)$ is given in Appendix B. As mentioned in the introduction, these
$SU(3)$ fields play a role only in the Chern Simons action. They can be set to zero elsewhere. The effective DBI action
is just a $U(2)$ non Abelian DBI action. 

When the branes are separated, and the non-Abelian fields are not excited,
this reduces to two decoupled Abelian DBI actions. In field theory language, we have a $U(1)\times SU(2)$ group broken to a $U(1)\times U(1)$. The Higgsing of the $SU(2)$ is by the adjoint field corresponding to the separation of the branes. This is just the Georgi-Glashow model.  $U(1)\times U(1)$ can be denoted  $U(1)_B\times U(1)_3$ as in (\ref{u1}). Thus
$A = A_B {\mathbf I} + A_3 t_3$.

We can factor out a 3-volume
$V_3$ of the three dimensional space of the boundary theory,  by writing $N_4 = n_4 {V_3\over R^3} \;$. $n_4$
is also dimensionless. In terms of these we get
\be 
S_{CS} = {N_c N_4} \int d t_E A_0(u_c) 
\equiv n_B \tilde T a_0(u_c)
\ee
where, integrating over the Euclidean time $t_E$ with
circumference $\beta$, we have
\[
n_B \tilde T 
= N_c n_4 {V_3\over R^3} \beta {R\over 2\pi \alpha'} 
= {N_c n_4 V_3 \beta \over R^2 2 \pi \alpha'} \; \; . 
\]
The instanton number $N_4$ is thus the number of $D4$ branes
and equivalently the number of baryons. The charge of a baryon
is taken to be $N_c$.
  
The action of $N_4$ $D4$ branes wrapped on $S^4$ is 
\[
S_{D4} = {1\over (2\pi)^4 l_s^5} \int d t_E d^4 \Omega \; 
e^{-\phi} \sqrt{-g_{00} g_{S^4}}
= { N_4 R^4 \Omega_4 \beta \over (2\pi)^4 l_s^5 g_s} \; u_c
\]
where we have integrated over the Euclidean time $t_E$ with
circumference $\beta$. From the $u_c$ dependence we see that
there is a downward force.
  
Using $N_4 = n_4 {V_3 \over R^3}$, $g_s l_s^3 = {R^3\over N_c}$,
and $\Omega_4 = {8 \pi^2 \over 3}$ we get for the D4-brane
action:
\be
S_{D4} = {N_c n_4 V_3 \Omega _4 \beta \over (2\pi)^4 R^2 l_s^2}
u_c = {n_B ~\tilde T \over 3} u_c   \; \; . 
\ee

\subsection{Solution of DBI action with source}
 
The first step is to find the background $D8$ brane
configuration. To this end we solve the equations of motion of
the $D8$ brane Dirac-Born-Infeld action with a background gauge field
on the $D8$ brane worldvolume.
\be \label{SSDBI}
S_{DBI}=-\tilde T  \int d \sigma~ 
{u^{5 \over 2} \over \sqrt{f}} \; \; [ u'^2 + u^3 f^2 \tau'^2 
- f (\partial_\sigma a_0)^2 ]^{1 \over 2} 
- n_B \tilde T a_0(u_c)
\ee
where $u' = \partial_\sigma u \;$, $\tau' = \partial_\sigma \tau
\;$, $n_B$ was defined above and gives the magnitude of the
charge to which $a_0$ couples, and $\tilde T={V_3 R^5 \Omega_4
\over l_s^9 g_s}$. Let
\[
\Pi_a = \frac { u^{5 \over 2} \; \sqrt{f} \; (\partial_\sigma
a_0) } { [ u'^2 + u^3 f^2 \tau'^2 - f (\partial_\sigma a_0)^2
]^{1 \over 2} } \; \; .
\]
Then the equation of motion for $a_0$ is 
\be
\partial_\sigma \; \Pi_a = {n_B} \; \delta (u-u_c) \; \; . 
\ee

Choose the coordinate $\sigma$ along the brane to be equal to
$u$, the bulk radial coordinate. Then $u' = 1 \;$ and $\tau' =
\partial_u \tau \;$.

\vspace{2ex}

{\bf Case 1: $n_B = 0$}

\vspace{2ex}

We then find in the absence of sources that $\Pi_a = d_0$, a
constant. Thus,
\be
\frac {u^5 \; f \; (\partial_u a_0)^2} 
{1 + u^3 f^2 \tau'^2 - f (\partial_u a_0)^2} = d_0^2
\ee
Solving for $\partial _u a_0$ we find
\be
(\partial _u a_0)^2= {d_0^2 \; (1 + u^3 f^2 {\tau '}^2 )
\over f \; (u^5 + d_0^2)}
\ee
One can plug this back into the action and evaluate the Legendre
transform of the Lagrangian density: $\tilde {\cal L}={\cal L} -
~ \tilde T \; \Pi_a \; \partial _u a_0 \;$ :
\be
\tilde {\cal L}(d_0) = \tilde T \; 
\sqrt{u^5 + d_0^2} \; \sqrt{{1 \over f} + u^3 f {\tau '}^2}
\ee

We can repeat this for $\tau' = \partial_u \tau $ and write the
Legendre transformed Lagrangian density ${\cal L}(c_0,d_0)$:
\be   \label{Legendrec0}
\tilde{\cal L} (c_0,d_0)= -\tilde T \; 
{\sqrt{D} \over u^{3 \over 2} f}
\ee
\be \label{tau'}
(\partial_u a_0)^2= {d_0^2 \; u^3 \over D} 
\; \; \; , \; \; \; \; 
(\partial_u \tau)^2 = {c_0^2 \over u^3 f^2 D}
\ee
where $D = u^8 f + d_0^2 u^3 f - c_0^2 \;$. When $D \;$ vanishes
(for some $u=u_0$), $\tau' = \infty$. This is the lowermost
point of the $D8-$brane and here the vertical force due to the
brane tension vanish because the brane is horizontal. The
horizontal forces cancel if we add a mirror $D8-$brane so that
we have a symmetric U--shaped configuration. In the Sakai
Sugimoto interpretation the second set of branes provide
fermions of the opposite chirality, and are referred to as
$\bar{D}8-$branes. The joining of these two sets of $D8-$branes
represents the breaking of chiral symmetry.

Naively the gauge field strength diverges at $u_0$. However this
is a ``coordinate singularity'' - the choice $\sigma = u$ is not
good at $u = u_0$ because $\sigma$ keeps increasing
monotonically as we proceed along the bottom of the U--shape,
whereas $u$ starts to decrease again. See  \ref{appkr1}.
In this appendix a more general analysis is done keeping the coordinate
$\sigma$ throughout. Thus $(\partial _u a_0)= { (a_0) _\sigma \over u _\sigma}$
and  $(\partial _\tau a_0)= {(a_0)_\sigma \over \tau _\sigma}$ can be calculated for
a given solution.\footnote{We use the notation $f_\sigma= {\partial f\over \partial \sigma}$} The former is singular at $u_0$ because $u_\sigma$ vanishes at $u=u_0$, whereas the latter is finite.
Calculating the latter can also be accomplished equivalently very simply by  
 choosing $\sigma = \tau$ and investigating the same 
solution. One finds:
\be
(\partial _ \tau a_0)^2= {d_0^2 u^3f\over d_0^2+u^5} \; \; . 
\ee
Note that $u_0$ is a solution of $D(u_0) = 0 \;$. So $u_0^5 +
d_0^2 = {c_0^2 \over (u_0^3 - u_{KK}^3)} \;$. This gives
\[
\partial _\tau a_0 |_{u=u_0}= {d_0^2 \over c_0^2}
\; (u_0^3 - u_{KK}^3)^2
\]
which is finite.

 In the absence of charges, continuity of flux
would require that the electric field continue in the same
direction along the other brane and reemerge on the
boundary. Thus $\partial _\sigma a_0$ is continuous. On the
other hand $\partial _\tau a_0$ changes sign because $\sigma = -
\tau$ on the mirror brane. Since the equations above fix only
the magnitude of $\partial _\tau a_0$, this is also a valid
solution. The conclusion is that it is possible to have $d_0\neq
0$ even when $n_B=0$. 
\vspace{2ex}

{\bf Case 2: $n_B \neq 0$}

\vspace{2ex}

In this case there is a jump in the value of $\Pi_a$ at $u =
u_c$. We will choose a solution where $\Pi_a = 0$ for $u < u_c$
and equal to $n_B$ for $u > u_c$.
\be
\partial_u \tau= \pm {c_0 \over u^{3 \over 2} f \; D} \; ,
~~~~~ u > u_c
\ee
where $D = u^8 f + n_B^2 u^3 f - c_0^2 \;$. $u_c$ is determined
by minimizing the action with respect to variations of $u_c$,
{\em subject to} the constraint that
\be
\int _{u_c} ^\infty du ~ \tau ' = {L\over 2}
\ee
$L$ is the distance between the $D8$ brane and the anti-$D8$
brane (assumed symmetric) at $u=\infty$. Note that the above
constraint implies that
\be   \label{Constraint}
\int _{u_c}^\infty du~ {d \tau '\over d u_c}-\tau '(u_c)=0
\ee
We work with $\tilde S (d) = -\tilde T \int _{u_c}^\infty du
\sqrt {u^5 + n_B^2 }\sqrt{ {1 \over f} + u^3 f \tau '^2}$ for
the D8 brane.  The variation of $u_c$ gives the upward force due
to the D8 brane tension. In order to get the downward force we
(following \cite{BLL, CDKLY}) use the action of a D4 brane
wrapped on $S^4 \;$. The answer evaluated above is ${n_B \tilde
T \over 3} u_c \;$. This gives a force of $n_B \tilde T \over 3
\;$. We do not need a horizontal force balance as the mirror
$\bar{D}8-$brane automatically provides the correct balancing
force.

One can do the variation keeping either $L$ fixed or $c_0$
fixed. The latter is easier, since we just work with the
Legendre transformed action:
\be   
- \tilde T \int _{u_c}^\infty {d u  \over u^{3 \over 2} f} \;
\sqrt{u^8 f + n_B^2 u^3 f - c_0^2 }
\ee
Varying $u_c$ just sets the integrand $\tilde{\cal L}(c_0, n_B)$
equal to $n_B \tilde T\over 3$, the force due to the
D4-brane. One can solve for $u_c$ in terms of $c_0, n_B$. The
result for $c_0=100, u_{KK}=1$ is plotted in the figure.

If one substitutes for $c_0$ in terms of $\tau '$ one can
manipulate the D8-brane variation into the form:
\[
\tilde T \; \; 
\underbrace{1 \over u^{3 \over 4} \sqrt f}_{\sqrt {g_{uu}}} 
\; \;
\underbrace{u^{-{3 \over 4}}\sqrt{u^8+u^3 n_B^2}}_{tension} 
\; \;
\underbrace{1 \over \sqrt{1 + u^3 f^2 \tau'^2}}_{sin ~\alpha} 
\; \; .
\]
In this form it is easy to see the physical interpretation. The
factor ($sin~\alpha$) gives the vertical component of the force.
One also finds $sin^2 \alpha_c = {n_B^2\over u_c^5 + n_B^2}$
where $\alpha_c$ is the angle with respect to the horizontal at
$u=u_c \;$.
 
\begin{figure}[htbp]
\begin{center}
\includegraphics{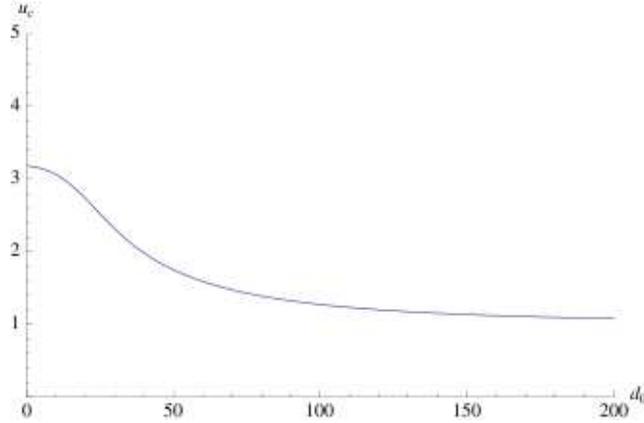}
\caption{$u_c$ versus $d_0 (= n_B)$ with $c_0=100$}
\label{default}
\end{center}
\end{figure}

\subsection{$SU(2)$ Breaking}
   
Our aim is to introduce a finite ``baryon number" corresponding
to the $U(1)$ subgroup of $SU(2)$ generated by $t_3 \;$. 
 When this is done there are configurations of the type shown in 
the second diagram in Figure 1 where the two branes separate. The branes are parallel
at large $u$, but separated by a finite distance. This Higgses the $SU(2)$ to $U(1)$
as in the Georgi-Glashow model.

The
question is whether there will be a condensate that
spontaneously breaks this $U(1)$ as happens in the BCS
case. Note that on the boundary this is a global symmetry, so in
this sense it is like superfluidity. However what is non trivial
is that a fermion bilinear must condense in order for this
symmetry to break. This is more like BCS than condensation of a
bosonic field.
      
$u_c$ has a (rather complicated) dependence on $d_0 (= n_B) \;$
and $c_0 \;$. Also, the angle of the brane depends on $n_B$. We
have to choose values so that $u_c$ has the same value for both
branes. Thus to begin with $c_1 = c_2 ; ~~ d_1 = d_2 \;$. We then
turn on $d_1 - d_2 \;$, but change $c_1, c_2$ such that $u_c$
remains the same. This can be done numerically.
      
Thus if we turn on a  charge corresponding to $t_3$, which
is the difference $n_{(3)} = n_{B1} - n_{B2} \;$, then we
expect that the two branes will subtend different angles at $u_c
\;$. Thus the situation is as shown in Figure (\ref{cusp}). For small
values of $n_B$ and the angle $\alpha_c \;$, $\alpha_{c1, d_1} -
\alpha_{c2, d_2} \propto n_{B1} - n_{B2} = n_{(3)} \;$.
      
When two branes meet at an angle, in general there is a
tachyonic mode. This mode corresponds to the tendency of the
branes to smoothen out the angle and straighten out - see Figure (\ref{tach})
.  In the presence of the electric field the situation is
somewhat different. The presence of the electric field implies
that there must be open strings which connect the two branes and
carry the electric flux. The tension of these open strings then
in principle can prevent the separation of the branes. In fact
a perturbative stability analysis shows that there is no
tachyon.

In general one expects the relation between the slope $\alpha_c$
and the electric field $n_B$ to be modified by the condensate.
The condensation corresponds to $\phi^{\pm}$ and in general also
$w^{\pm}$.

The background configuration breaks $SU(2)$ . The Non Abelian DBI
action \cite{Myers} is well defined as long as the background is
diagonal, which means it lies in the $U(1) \times U(1)$ subgroup
of $U(2)$. Otherwise there are ambiguities in the definition of
the determinant . There is the symmetrized trace prescription
that is often used, though in general it is known that there are
modifications at the $O({\alpha '})^6$\cite{SW, KS}.

When the $SU(2)$ fields are small 
one can study this problem in the Yang-Mills approximation to
the Dirac-Born Infeld. In Section 3 we give an exact solution of the Yang-Mills
equations in flat space. This solution is not the most
general. It has $\phi^+ = 0 \;$, although $w^{+} \neq 0 \; $. As
a consequence the asymptotic slope is still fixed by the
electric field.

In the presence of the D4 D8 background and the supersymmetry
breaking metric function $f(u) \;$, we find an approximate
solution of the Yang-Mills equations by neglecting some terms in
the equation of motion. We find that at least when $u_c >>
R_{AdS}$ the neglected terms are small and we expect our
solution to be a good approximation.  This is described in
Section 4. To do better than this one needs a numerical
solution. We give some exact numerical solutions in Section 4.
The solutions where the $SU(2)$ fields are small a posteriori
justify the Yang-Mills approximation and are thus very close to
the exact solutions of the DBI action.

The conclusion is thus that as soon as we turn on a 
charge (corresponding to $t_3$) we have a solution where there
is a charge violating condensate. This must be a
reflection of the fact that from the boundary theory viewpoint,
as soon as we have a Fermi surface (i.e. a non zero chemical
potential) there is an instability towards a BCS phase.
\footnote{ As mentioned earlier this conclusion is contingent on
the absence of other instabilities in the boundary that give
rise to the same kind of symmetry breaking.}

As a caveat one should note that this analysis does not take
into account the tachyon condensation that represents the
fermion mass generation. However this should not affect the BCS
instability, which only requires gapless excitation above the
Fermi surface. The existence of this is not affected by the
fermions being massive.

In the next section we analyze the stability of this solution in
flat space using Yang-Mills as an approximation to Dirac-Born- Infeld.

\section{The Yang-Mills Approximation}

We will assume that the fields on the $D8$ brane depend only on
$(t, u) \;$, thus keeping translational invariance in $(x^1,
x^2, x^3) \;$. The $S^4$ coordinate dependence is also not
assumed. The $D8$ brane has only one transverse direction,
$\tau$, which we call $\phi$ here. The index $\mu$ on the
Yang-Mills gauge field $A_\mu$ has nine values. However, we set
$A_\mu = 0$ along the four directions of the $S^4 \;$. For the
stability analysis we concentrate on $A_0, A_u$ and $\phi$. For
other purposes, such as calculating correlators of the boundary
theory, we also use $A_{x^1}$.

We define the $SU(2)$ generators $(t_+, t_-, t_3) \;$:
\be
t_+ = \left( 
\begin{array}{cc}
0 & 1 \\
0 & 0 
\end{array} \right) 
~~~~~~ t_- = \left( 
\begin{array}{cc}
0 & 0 \\
1 & 0 
\end{array} \right)
~~~~~~ t_3 = \frac{1}{2} \; \left( 
\begin{array}{cc}
1 & 0 \\
0 & - 1 
\end{array} \right)
\ee
which satisfy the $SU(2)$ algebra 
\[
[t_3, t_{\pm}] = \pm \; t_{\pm} , ~~~~
[t_+, t_-] = 2 \; t_3 ~~~~
\{t_+, t_-\} = I \; \;. 
\]
If we write $A = A^+ t_+ + A^- t_- + A^3 t_3 \;$ then $Tr(AA) =
\hf [(A^3)^2 + 4 A^+ A^-] \;$. 

In this basis we define the Yang-Mills field $A_\mu = A^+_\mu
t_+ + A^-_\mu t_- + A^3_\mu t_3 \;$ and, similarly, an adjoint
scalar field $\phi = \phi^+ t_+ + \phi^- t_- + \phi^3 t_3 \;$
where $\phi^3$ is real and $\phi^\pm$ are complex conjugates of
each other. The covariant derivative is defined by
\[
D_\mu \phi = \partial _\mu \phi + i [A_\mu, \phi]
\] 
and the field strength by 
\[ 
F_{\mu \nu} = \partial_\mu A_\nu - \partial_\nu A_\mu
+ i [A_\mu, A_\nu] \; \;. 
\] 
We can write the Yang-Mills and scalar field action as
\be
S = - {1\over 2} Tr[F_{\mu \nu} F^{\mu \nu}] 
- Tr[D_\mu \phi D^\mu \phi] \; \; . 
\ee

\subsection{Solutions of the Equations of Motion}\label{3.1} 

We first study the case where the background space time is
flat. Later, we will study the case where the background
space time is curved.

\subsubsection{Solution with $A_\mu  = 0$}

The simplest non trivial solution is of course the Higgs phase
where $\phi^3 = const$. The resulting unbroken symmetry is just
$U(1) \;$. In the brane language this corresponds to a constant
separation of the $D8$ branes in the $\tau$ direction.
 
The configuration of intersecting $D8$ branes corresponds, in
the Yang-Mills description to having $\phi = q u t_3 \;$ where
$q$ is the slope. Clearly this satisfies the equations of motion
\[
D^\mu D_\mu \phi = 0 \; \; , \; \; \; 
D^\mu F_{\mu \nu} = [\phi, D_\nu \phi]
\]
where $D_\mu (*) = \partial _\mu (*) + i [A_\mu, (*)] \;$. For
the given configuration $D_u \phi$ is constant and $A_\mu$ is
zero, so it clearly satisfies the equations of motion.

When we embed $D8$ branes in $D4$ brane background, as the
discussion in the previous section shows, there are solutions
that interpolate between these two solutions. However this
requires a non zero $A_0$.

\subsubsection{Solutions with $A_0 \neq 0$}

We can have a configuration where $A_0^3 = - E u \;$, in
addition to the above $\phi \;$. This corresponds to a constant
electric field $E$ and also clearly satisfies the equations of
motion. It is this solution (near $u = u_c$) that interpolates
with the $\phi = const$ solution near $u = \infty \;$, in the
$D4$ background.

{\bf Solution I:} 

We choose a gauge where $A_0^\pm = A_u^3 = 0 \;$. By imposing
this everywhere including the boundary, we are freezing the
Goldstone modes (pions) \cite{SS}. We consider time independent
configurations and also set the charged scaler to zero. We list
below the various equations, after the choice of gauge :
\footnote{ Note that gauge invariance of the action implies the
following relation between the equations: $D_\mu {\partial \cal
L \over \partial A_\mu} +i e [\phi,{\partial \cal L \over
\partial \phi}] = 0 \;$. This means that after the gauge choice,
although we seem to have an over determined system of equations,
the equations are not all independent. \label{fn7}}

The equations are:
\be \label{fs1}
\du^2 A^3_0 - 4 w^+ w^- A^3_0 = 0  
\ee
\be\label{fs2}
\du^2 \phi^3 - 4 w^+ w^- \phi^3 = 0 
\ee
\be \label{fs3}
\du (w^+ A^3_0) + w^+ \du A^3_0 = 0 
\ee
\be\label{fs4}
((A^3_0)^2 - (\phi^3)^2) \; w^+ = 0
\ee
\be\label{fs5}
\du (w^+ \phi_3) + w^+ \du \phi_3 = 0 
\ee
Eqn (\ref{fs3}) implies that $(A_0^3)^2w^+=c$, where is $c$ is a
constant. Clearly $\phi^3=A_0^3$ is required too. Finally
plugging into (\ref{fs1}), Gauss law, we can integrate to get
\be
A_0^3(u)=\phi^3(u)=\sqrt{{du^2\over 4}+{4c\over
d}};~~~~~~w^+=w^-={c\over {du^2\over 4}+{4c\over d}}
\ee 
The D8 brane has the expected profile - the intersection region
is smoothed out, but it does not straighten out fully, and there
is a condensate of $w^\pm$ representing open strings connecting
the two D8 branes. This solution was also studied in~\cite{HW}.

{\bf Solution II:}

A solution can also be obtained if there are two adjoint scalars
$\phi_1$ and $\phi_2 \;$, with the action given by
\be
S = - {1\over 2} Tr[F_{\mu \nu} F^{\mu \nu}] 
- Tr[D_\mu \phi_1 D^\mu \phi_1] 
- Tr[D_\mu \phi_2 D^\mu \phi_2] 
+ Tr[\phi_1, \phi_2]^2 \; \; . 
\ee
We define 
\begin{eqnarray*}
\chi & = & \chi^3 t_3 + \chi^+ t_+ + \chi^- t_-
\equiv \phi _1  + i \phi_2 \\
& = & (\phi_1^3 + i \phi_2^3) t_3 
+ (\phi_1^+ + i \phi_2^+) t_+ 
+ (\phi_1^- + i \phi_2^-) t_- 
\end{eqnarray*}
and, in an obvious way, 
\begin{eqnarray*}
\chi^* & = & (\chi^3)^* t_3 + (\chi^+)^* t_+ + (\chi^-)^* t_-
\equiv \phi _1  - i \phi_2 \\
& = & (\phi_1^3 - i \phi_2^3) t_3 
+ (\phi_1^+ - i \phi_2^+) t_+ 
+ (\phi_1^- - i \phi_2^-) t_- \; \; , 
\end{eqnarray*}
and, finally, 
\[
\bar \chi \equiv (\chi^*)^T
= (\chi ^3)^* t_3 + (\chi^+)^* t_- + (\chi^-)^* t_+ \; \; . 
\] 
Thus $\bar \chi^+ = (\chi^-)^* \;$, $ \; \bar\chi ^- = (\chi
^+)^* \;$, and $ \; \bar \chi^3 = (\chi^3)^* \;$. 
 
Of course if we set $\chi = 0$ we have the same solution as
before. Furthermore, if we set one of the scalar fields, such as
$\phi_1 = 0 \;$, then the system is again the earlier one and we
only have an analytic solution where $\phi_2$ is also zero. So
we try to set a different set to zero. One can try for instance
to set $\phi_1^3 = 0 = \phi_2^\pm \;$

The equations and details of the solution are given in the
Appendix {\bf F}.

When all the dust settles we find that $\chi^+$ can be
nonzero: either a constant or linear in $u$. But this changes
fairly dramatically the behavior of $A_0^3$ - it becomes
exponential rather than linear. If we take $\chi^+ = \chi_0 =
constant$ then $A_0^3$ can be solved for in closed form: One
finds using the same methods as earlier
\be
A_0^3(u) = \sqrt{a \chi_0} \; cosh ~(\chi_0 \; u)
\ee 
Note that $\du A^3_0 (0) = 0 \;$, and thus the electric field
vanishes at $u = 0$ as required by symmetry.


\subsection{Stability}\label{stability}

In this section we study the stability of the solutions given in
the last section. The stability of the solution with linear
$\phi^3$ has been analyzed by \cite{N, HN, HW, EL} and as
mentioned earlier a tachyon was found. This tachyonic instability
is illustrated in Figure \ref{tach}.

The instability would continue till the initially intersecting
branes straighten out completely, if we take this linearized
picture seriously. However what actually happens is more
complex. The condensing tachyon is a charged field. Once there
are charges one has to take into account electric fields. The
situation when there is an electric field present is described
in Figure \ref{E}. Since the electric fields point in opposite
directions on the two branes (we are considering the $U(1)$
generated by $t_3$ of $SU(2)$), as the branes separate open
strings need to be formed to carry the electric flux. These open
strings between the separating branes have tension, oppose
stretching, and do not want to stretch beyond a point. This
stabilizes the configuration.  

This intuitive picture is consistent with the linearized
stability analysis given below: One finds that $(mass)^2 = {(q^2
- E^2)^{3 \over 2} \over q^2} \; (2 n + 1) \; $ when $q^2 > E^2
> 0 \;$, and $ (mass)^2 = (2 n - 1) \; \vert q \vert$ when $E =
0 \;$, with $n = 0, 1, 2, \cdots \;$. Thus, the lowest $(mass)^2
\propto (q^2-E^2)^{3\over 2} > 0$ when $E \ne 0 \;$, and the
lowest $(mass)^2 < 0$ when $E = 0 \;$.

The $N_f = 2 \;$ number of $D8$ branes in our case are in a
background space curved by the $D4$ branes. Nevertheless near
the cusp at $u = u_c \;$, where these two $D8$ branes intersect,
the local physics can be studied using the flat space model.
This is what is done below. 

The equations for small fluctuations is derived in the 
Appendix {\bf G}. 

Let us assume $\phi^+ (t,u) = \phi(u) \; e^{-i m t} \;$ and
$w^+(t,u) = i w(u) \; e^{-i m t} \;$. Using $\phi^3 = q u$ and
$A^3_0 = - E u \;$, we then get
\be \label{du2phi}
\partial_u^2 \phi + q \; (2 + u \partial_u) w
+ (m + E u)^2 \phi  =  0 
\ee
\be \label{duw}
(q^2 u^2 -(m+Eu)^2) w=q(1-u\pa_u)\phi
\ee
Define
\[
b^2 = q^2 - E^2 \; \; , \; \; \; 
c = \frac{m^2 q^2}{b^2}  \; \; , \; \; \; 
\tilde{u} = u - \frac{m E}{b^2} 
\]
and the functions $P(u) \;$ and $Q(u) \;$ by
\[
P(u) = \frac{2 m q^2}{(m + E u) \; Q(u)} 
\; \; \; , \; \; \; \; \;
Q(u) = q^2 u^2 - (m + E u)^2 = b^2 \tilde{u}^2 - c 
\; \; . 
\]
Then using (\ref{duw}) the equation (\ref{du2phi}) becomes
\be\label{44}
\du^2 \phi - u P \; \du \phi - (Q - P) \; \phi 
\; = \; 0 \; \; .
\ee

For large $u \;$, we have $u P \sim 0 \;$, $P \sim 0 \;$, and $Q
\sim b^2 u^2 \;$. So, asymptotically for large $u \;$, this is a
Schr\"{o}dinger equation for harmonic oscillator. Let $\phi = e^{-
\frac {b} {2} \tilde{u}^2} F \;$ where $b = \sqrt{q^2 - E^2} \;
> 0 \;$ and $\tilde{u} \;$ has been defined earlier. We then
have
\be 
\du^2 F - (2 b \tilde{u} + u P) \; \du F 
+ (c - b + (1 + b u \tilde{u}) P) \; F \; = \; 0 \; \; .  
\ee
Letting $F(u) = u^n \;$, where $n \ge 0 \;$ is an integer, and
noting that $P \sim \frac{1}{u^3} \;$ for large $u \;$, we get
from the coefficient of $u^n$ the eigenvalue condition: $c = (2
n + 1) \; b \;$, which shows that $m^2 = {(q^2 - E^2)^{3 \over
2} \over q^2} \; (2 n + 1) \; $ is always non-negative. When
$q^2 = E^2 \;$, the spectrum is gapless.

This is to be contrasted with the situation where $E = 0 \;$.
Observe that $P \sim \frac{2 m q^2}{b^2 E} \; \frac{1}{u^3} \;$
for large $u \;$ if $E \ne 0 \;$, irrespective of how small
$\vert E \vert$ is. But, if $E = 0 $ exactly then $P \sim
\frac{2 q^2} {b^2} \; \frac{1}{u^2} \;$ for large $u \;$. This
results in the $(b u \tilde{u} P F)-$term contributing to the
eigenvalue condition which is now: $c = (2 n + 1) \; b - \frac{2
q^2}{b} $ which gives $ m^2 = (2 n - 1) \; \vert q \vert$ since
$b = \vert q \vert$ and $c = m^2 $ when $E = 0 \;$. This
condition reveals the presence of a tachyonic mode when $E = 0
\;$.

The new solution with $w^\pm \neq 0$, described in the last
section has the same asymptotics as the above solution, hence we
do not expect a tachyon. Nevertheless, since $q=E$, there are
massless modes and so we expect that there are continuous
deformations. This is certainly true since the solutions are
characterized by integration constants $(c, d) \;$ which are
free parameters and can be changed continuously.

Finally, the fact that the negative $(mass)^2$ is proportional
to the angle, which in turn, is proportional to the number
density, suggests that the critical temperature at which
superconductivity is lost is proportional to a power of the
number density. This is different from the usual BCS relation
and is closer to the strong coupling BCS result given in the
Introduction.

\section{Solutions in Curved Space Time}

In this section we give the curved space counterparts of the
flat space equations given in Section {\bf \ref{3.1}}. We also
attempt to find a solution that is the counterpart of the one
presented there. Our aim will be to establish in a qualitative
way that the flat space solutions presented in Section {\bf
\ref{3.1}} have a generalization to the $D4 D8$ background. A
quantitative (numerical) study of these equations is not
attempted in this paper.

\subsection{Equations of Motion}

Our starting point is the Dirac-Born-Infeld action, with $\sigma = u
\;$, 
\be 
S_{DBI} = -\tilde T \int d^4 x ~ d u~ 
{u^{5 \over 2} \over \sqrt{f}} \; \; [ 1 + u^3 f^2 \tau'^2 
- f (\partial_u a_0)^2 ]^{1 \over 2} 
- n_B \tilde T a_0(u_c)
\ee
For the moment we ignore the source term. We have 
\be
[1 + u^{3} f^2 \tau'^2 - f (\du a_0)^2]^\hf \approx
1 + \hf u^{3} f^2 \tau'^2 - \hf f (\du a_0)^2 \; \; . 
\ee
The last two terms are nothing but $\hf g^{u u} g_{\tau \tau}
(\partial_u \tau)^2 + \hf g^{0 0} g^{u u} f_{0 u}^2 $ which are
the world volume scalar kinetic term $g_{\tau \tau} \;
\partial_\mu \tau \partial^\mu \tau \;$ and the Maxwell term
$f_{\mu \nu} f^{\mu \nu} \;$ written in curved space-time. The
factor $g_{\tau \tau}$ is a component of the background
space--time metric in the transverse direction.

For the case of two D8 branes, we need the non-Abelian
Dirac-Born-Infeld action \cite{Myers}. However this is not very well
defined although there are prescriptions that are known to be
consistent up to some order \cite{SW, KS}. We will bypass this
complication by expanding the Dirac-Born-Infeld action as was done
above and using the Yang-Mills approximation for the non Abelian
gauge field $A_\mu \;$ and keeping the Dirac-Born-Infeld structure for
the $U(1)$ gauge field $a_\mu \;$. In order to get the correct
normalization of the the commutator term $[A_0,\tau]^2$ we will
start with a $U(1)$ theory that has time derivatives of $\tau
\;$, covariantize and then set the time derivative to zero.
See \ref{appbs1} and \ref{appkr2} also. 

Thus our starting point is:
\[
[(g_{00} + g_{\tau \tau} \dot{\tau}^2) (g_{uu}+ g_{\tau
\tau}\tau'^2) + f_{0u}^2]^\hf
\]
\[
=[ g_{00}g_{uu}(1+ g^{00} g_{\tau\tau}{\dot \tau}^2)(1+
g^{uu}g_{\tau \tau}{\tau '}^2))+ f_{0u}^2]^\hf
\]
\[
= [g_{00}g_{uu}((1+ g^{00} g_{\tau\tau}{\dot \tau}^2)(1+
g^{uu}g_{\tau \tau}{\tau '}^2)+ g^{00}g^{uu}f_{0u}^2)]^\hf
\]
\[
=[g_{00}g_{uu}((1+ g^{uu}g_{\tau \tau}{\tau '}^2+
g^{00}g^{uu}f_{0u}^2) + (1+ g^{uu}g_{\tau \tau}{\tau '}^2)g^{00}
g_{\tau\tau}{\dot \tau}^2)]^\hf
\]
Covariantize now $\dot \tau$ and set the time derivative to zero
: $\dot \tau \rightarrow D_0 \tau \rightarrow i [A_0, \tau] \;$.
This gives:
\be
S_{DBI} = - \tilde T \int d^4 x ~ d u ~ 
{u^{5 \over 2} \over \sqrt{f}} \; \sqrt{X} \; 
- \; n_B^{U(1)} a_0(u_c) \; 
- \; \vec n_B^{SU(2)} \cdot \vec A_0(u_c)
\ee
where $X$ is given, denoting the $SU(2)$ part of $\tau$ by $\phi
\;$, by
\[
X \equiv 
1 + g^{u u} g_{\tau \tau} [\tau ^{'2} + 2 Tr({D_u \phi})^2]
+ \; g^{00} g^{uu} [(\partial_u a_0)^2 + 2 \; Tr(F_{0 u})^2]
\]
\[
+ (1 + g^{uu} g_{\tau \tau} \tau'^2) g^{0 0} g_{\tau \tau} 
\; 2 Tr( i \; [A_0,\phi])^2
\]
Note that the factor of 2 for $SU(2)$ traces is because the
$t_3$ generator is defined to be $\hf \tau_3$ where $\tau_3$ is
the Pauli matrix.

For the $SU(2)$ part, the covariant derivative $D_\mu * =
\partial_\mu * + i [A_\mu, * ]$ has been introduced in place of
the ordinary derivative, and $F_{\mu \nu} = \partial_\mu A_\nu -
\partial_\nu A_\mu + i [A_\mu, A_\nu]$ is the non-Abelian field
strength. Had the $SU(2)$ fields been entirely in the diagonal
$t_3$ direction, this action would have been exact. This was
discussed in Section 1. To the extant that there are off
diagonal terms, this action is not correct. It can also be shown
that if the off diagonal terms are entirely in the antisymmetric
$F_{0 u}$ part, then the symmetrized trace prescription gives
exactly this action. (A proof is given in the \ref{appbs1}).
Since the symmetrized trace prescription is itself known to be
correct only up to $O((\alpha ')^6)$ we will not belabor this
point here. However we assume for the moment that the off
diagonal terms are small and work with this action and expand
the $SU(2)$ part in the square root keeping the background
$U(1)$ inside the square root.

Defining $\Delta = 1 + g_{\tau \tau} g^{uu}\tau^{'2} + g^{00}
g^{uu} (\du a_0)^2 \;$, we have
\[
S_{DBI} \approx - \tilde T \int d^4x ~ du ~ 
{u^{5 \over 2} \over \sqrt{f}} \; \sqrt{\Delta} \; 
[1 + {1 \over 2 \Delta} \; 
[ g^{u u} g_{\tau \tau} \; 2 Tr(D_u \phi)^2
\]
\[ 
+ (1 + g^{uu} g^{00} \tau'^2) g^{0 0} g_{\tau \tau} \; 
2 Tr( i [A_0,\phi])^2
+ g^{00} g^{uu} \; 2 Tr(F_{0 u})^2 ] \; ]
\]
\[
= \tilde T \int d^4x ~ du ~ 
[{u^{5 \over 2} \over \sqrt{f}} \; \sqrt{\Delta} \;
+ \underbrace{u^{5 \over 2} \over 
\sqrt{f \; \Delta}}_{{\rm ``} \; \sqrt {-g} \; "}  \; 
[g^{u u} g_{\tau \tau} Tr(D_u \phi)^2
\]
\be\label{49}
+ \; (1 + g^{uu} g^{00} \tau'^2) g^{0 0} g_{\tau \tau} 
Tr( i [A_0, \phi])^2 + g^{00}g^{uu} Tr(F_{0 u})^2 \; ] \; ] 
\; \; . 
\ee
The pre factor multiplying the Yang-Mills action has been
denoted symbolically as $\sqrt{-g}$.

Note that if we had $g^{\tau \tau}$ instead of $g_{\tau \tau}$,
then in the limit $f=1$ the scalar could have been thought of as
another vector and the action would be the same as the flat
space action up to an overall factor. It is easy to see that in
this case the flat space solution given in the last section can
be generalized very directly to curved space.

We are interested in static and translation invariant (in the
$(x^1, x^2, x^3)$ directions) solutions of the equations of
motion. The fields thus depend only on $u$. The equations are
the following. These equations assume that $\tau' = 0 \;$,
i.e. neglect the bending of the D8 branes.
\[ 
{\delta S \over \delta A^3_0} = 
- \du \; ( \rg g^{0 0} g^{u u}\du A^3_0 )
+ 4 \; \rg g^{0 0} \; ( g^{u u} w^+ w^- 
+ g_{\tau \tau} \phi^+ \phi^- ) \; A^3_0
\]
\[
{\delta S \over \delta A^3_1} = 2 i \; \rg g^{u u} g_{\tau \tau}
\; ( \phi^- D_u \phi^+ - \phi^+ D_u \phi^- )
\]
\[ 
{\delta S \over \delta \phi^3} = 
\du (\rg g^{u u} g_{\tau \tau} D_u \phi^3) 
- 2 i \; \rg g^{u u} g_{\tau \tau} \; 
( w^- D_u \phi^+ - w^+ D_u \phi^- )
\]
\[ 
{\delta S \over \delta A^-_0} = 
2 i \; \du \; ( \rg g^{0 0} g^{u u} w^+ A^3_0 ) 
+ 2 i \; \rg g^{0 0} g^{u u} w^+ \du A^3_0
- 2 \rg g^{0 0} g_{\tau \tau} A^3_0 \phi^3 \phi^+
\]
\[ 
{\delta S \over \delta A^-_1} = 
2 i \; \rg g^{u u} g_{\tau \tau} \;
( \phi^+ D_u \phi^3 - \phi^3 D_u \phi^+ )
- 2 \rg g^{0 0} g^{u u} \; (A^3_0)^2 w^+ 
\]
\be\label{50}
{\delta S \over \delta \phi^-} = 
2 \du \; ( \rg g^{u u} g_{\tau \tau} D_u \phi^+ )
- 2 i \; \rg g^{u u} g_{\tau \tau} w^+ D_u \phi^3
- 2 \rg g^{0 0} g_{\tau \tau} (A^3_0)^2 \phi^+
\ee
where 
\[
D_u \phi^3 = \du \phi^3 - 2i \; (w^- \phi^+ - w^+ \phi^-)
\; \; \; , \; \; \; \; 
D_u \phi^+ = \du \phi^+ - i w^+ \phi^3 \; \; .
\]   
Here, we note the following.  Let $\phi^\pm = \phi \; e^{\pm i
\theta} \;$ and $w^\pm = w \; e^{\pm i \Omega} \;$ where $\phi$
and $w$ are real. It then follows from ${\delta S \over \delta
A^3_1}$ and ${\delta S \over \delta A^-_1}$ equations above that
$\Omega = \theta + \frac{\pi}{2} = constant \;$.

As mentioned above if we replace $g_{\tau\tau}$ by $g^{\tau
\tau}$ and set $f(u)=1$ (which is a good approximation if
$u>>u_{KK}$), then one can set $\phi^+=0$ as in flat space and
recover the same solution. Thus one would get $\rg \; (A^3_0)^2
w^+ = \rg \; (\phi^3)^2 w^+=const$, which except for the factor
$\rg$ is exactly as in flat space. But $g_{\tau \tau} = u^{3
\over 2}$, so clearly one cannot do this. However this suggests
a change of variables: Introduce auxiliary fields $\psi^a$
satisfying: $g_{\tau \tau} D_u \phi^a = g^{\tau \tau}D_u \psi^a
\;$. In all the equations where $\phi^a$ enter linearly, this
change gets rid of $\phi^a$ and those equations look exactly
like the flat space case. However where $\phi^a$ enter
quadratically, one ends up with both $\phi^a$ and $ \psi^a
\;$. This happens in the gauge field variations. However, since
$D_u \phi^a = (g^{\tau \tau})^2 \; D_u \psi^a = u^{- 3} \; D_u
\psi^a \;$, one expects that $D_u \phi^a$ are very small
compared to $D_u \psi^a \;$ for large $u \;$. It is plausible
that terms involving $\phi^a$ are much smaller than the other
terms, at least for very large $u \;$. Thus our strategy will be
to set these terms to zero as the zeroth approximation. The
equations of motion can now be solved analytically for $\psi^a
\;$. In terms of $\psi^a$ one can solve for $\phi^a \;$. In the
case where we set $\Delta = 1 \;$, this can in fact be done
analytically, though only a numerical solution is possible in
general. One can now verify whether there is some range of
parameters where the neglected terms are really small, so that
we have a self consistent procedure. We do find for very large
$u \approx 1000$ and specific values of the parameters, the
correction terms are a few percent of the terms we keep.
   
We give the substituted equations below. The terms involving
$\phi^a \;$ which are supposedly small are in boldface.

\be   \label{A03} 
{\delta S \over \delta A^3_0} = - \du \; (\rg \du A^3_0)
+ 4 \rg A^3_0 w^+ w^- + \boldsymbol {4 \rg \phi^-\phi^+ A^3_0}
\ee
\be   \label{a3}
{\delta S \over \delta A^3_1} = 
2 i \; \rg \; ( \phi^- \du \psi^+ - \phi^+ \du \psi^- )
+ 2 \rg \; (\phi^- w^+ + \phi^+ w^-) \;\psi^3 
\ee
\[ 
{\delta S \over \delta \phi^3} = \du \; (\rg \du \psi^3)
- 2 i \du \; (\rg (\psi^+ w^- - \psi^- w^+) )
\]
\be   \label{tau3}
- 2 i \; \rg \; (w^- \du \psi^+ - w^+ \du \psi^-)
- 4 \; \rg \; w^+ w^- \; \psi^3 
\ee
\be    \label{w0}
{\delta S \over \delta A^-_{0}} = 
2 i \; \du \; (\rg w^+ A^3_0) + 2 i \; \rg \; w^+ \du A^3_0
- \boldsymbol{ 2 \; \rg \; A^3_0 \phi^3 \phi^+}
\ee
\[ 
{\delta S \over \delta A^-_1} = 
- 2 \; \rg \; ( (A^3_0)^2 - \phi^3 \psi^3) \; w^+ 
+ \boldsymbol{ 2 i \rg ( \phi^+ \du \psi^3 - \phi^3 \du \psi^+)}
\]
\be   \label{w-}
+ \boldsymbol{ 4 \rg \phi^+ (\psi^+ w^- -\psi^- w^+) }
\ee
\[ 
{\delta S \over \delta \phi^-} = 
2 \; \du \; (\rg \; \du \psi^+) 
- 2 i \; \du \; (\rg \; w^+ \psi^3)
- 2 i \; \rg \; w^+ \du \psi^3  
\]
\be   \label{tau-}
- 4 \; \rg \; (w^- \psi^+ - w^+ \psi^-) \; w^+
- \boldsymbol{2\rg (A^3_0)^2 \phi^+}
\ee

Equation (\ref{a3}) can be satisfied identically by the reality
conditions: $\phi^\pm = \pm \; i \; \phi(u) \;$, $\; \psi^\pm =
\pm \; i \; \psi(u)$ and $w^\pm = w(u) \;$. The following
configurations solve the equations (\ref{A03}, \ref{tau3},
\ref{w0}, \ref{tau-}), when the terms in bold are neglected:
\[
A^3_0 = \sqrt{{D_A G(u)^2\over 2} + {2 C_A^2 \over D_A}}
\; \; \; , \; \; \; \; 
\rg (A^3_0)^2 \; w(u) = C_A 
\]
\[
\psi^3 = \sqrt{{D_3 G(u)^2\over 2} + {2 C_3^2\over D_3}}
\; \; \; , \; \; \; \; 
\rg (\psi^3)^2 \; w(u)=C_3
\]
\be
\psi = \sqrt{{D_+ G(u)^2\over 2} + {2 C_+^2\over D_+}}
\; \; \; , \; \; \; \; 
\rg (\psi^+)^2 \; w(u)=C_+
\ee
where $G(u) - G(u_c) = \int^u_{u_c} d u' \; {2 \over \sqrt {-
g(u')}} \; \;$. The above equations imply that
\be 
{D_A\over C_A}={D_3\over C_3}={D_+\over C_+} \; \; .
\ee
Given $\psi^a$ one can solve for $\phi^a \;$. There are some
constants of integration that can be fixed by choosing boundary
conditions for $\phi^3, \phi^+ \;$.

If equation (\ref{w-}) is to be satisfied by a finite $A^3_0 \;$
then $\phi^3$ cannot be negligibly small. Thus we choose the
boundary condition on $\phi^3$ so that (\ref{w-}) is satisfied
with finite $A^3_0 \;$. This implies that $\phi^3$ cannot be
very small.  It is also found that $\phi^3 \psi^3$ has
approximately the same functional form as $(A^3_0)^2$ only for
large $u \;$. So we restrict ourselves to large $u >100 \;$. We
also choose therefore to set $C_+ = 0 $ ($\implies \psi^+ = 0 $)
so that terms involving $\phi^3 \psi^+ $ are negligible.

Finally the various constants of integration have to be adjusted
so that the terms in bold are smaller than the others in each
equation.  For example, with the boundary being at $u = 1000 \;$,
$C_A = 10^{-7}; ~~ C_3 = 10^{-6}; ~~ D_A = 10^{-4}; ~~ C_+ = 0;
\; \; \phi^3 (1000) \simeq 4.47 * 10^{-6}; ~~~
\phi^+(1000) = 0 \;$. We also set $\Delta = 1$ as an
approximation.

Thus for these values, as an example, (\ref{w-}) evaluated at $u
= 900$ is $\simeq -3.97 * 10^{-12}$, and a typical large term in
the equation $(\rg (A^3_0)^2 w^+ )\vert_{u = 900} \simeq 5.81 *
10^{-7} \;$. The same quantities evaluated at $u = 100$ are
$\simeq - 3.02 * 10^{-12} $ and $1.0 * 10^{-7}$ respectively.
This shows that the equations are satisfied to a very good
accuracy.

As mentioned in the beginning of this section, all the
restrictions have to do with justifying perturbation theory
around an analytic solution. In order to get a more general
solution one has to resort to numerical methods. 

Some exact numerical solutions to the equations (excluding
(\ref{w-})) are given below along with the analytic
approximations.

Figure 9 and 10 give a comparison of some of the analytical and
numerical solutions. As expected the agreement is very good for
large $u$.

\begin{figure}[htbp]
\begin{center}
\includegraphics[scale=.8]{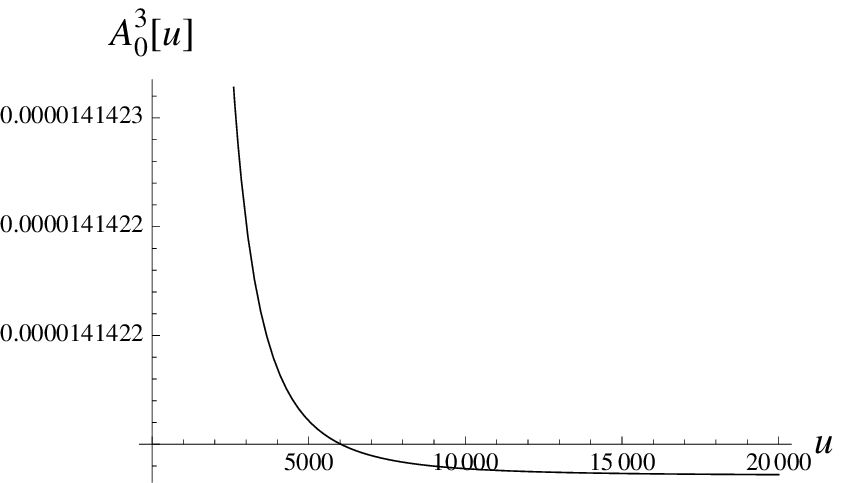}
\includegraphics[scale=.8]{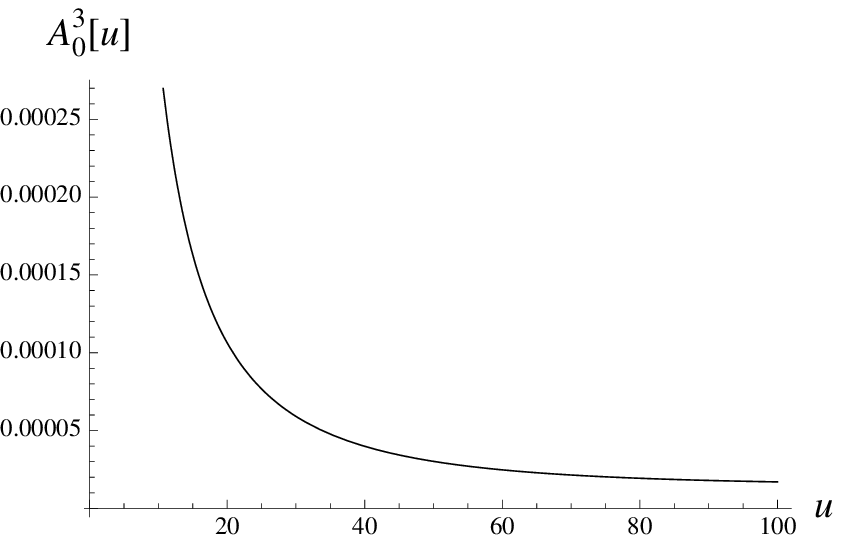}
\caption{Plot of $A_{0}^3$- numerical. The second plot expands
the small $u$ region.}
\label{figA03}
\end{center}
\end{figure}

\begin{figure}[htbp]
\begin{center}
\includegraphics[scale=.8]{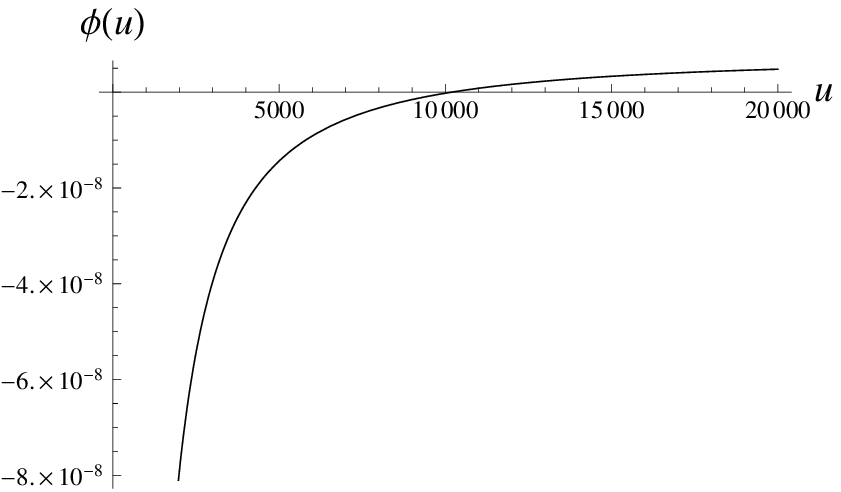}
\includegraphics[scale=.8]{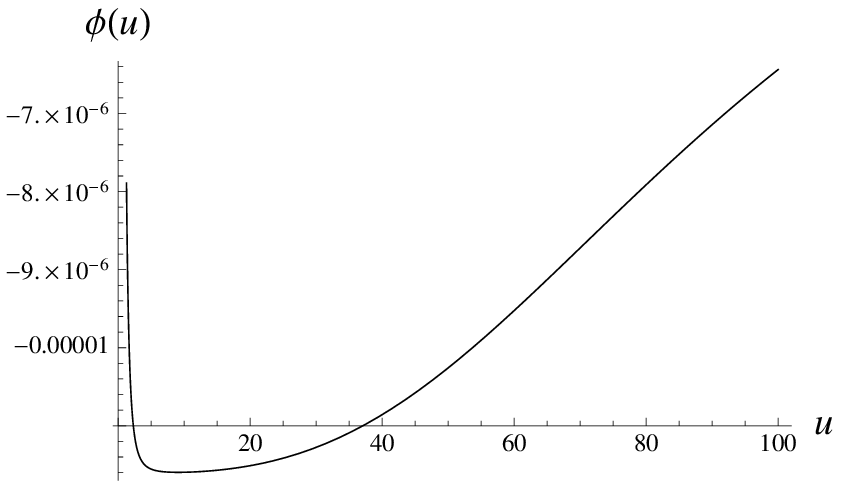}
\caption{Plot of the charged scalar field - numerical. 
The second plot expands the small $u$ region.}
\label{t}
\end{center}
\end{figure}

\begin{figure}[htbp]
\begin{center}
\includegraphics[scale=.8]{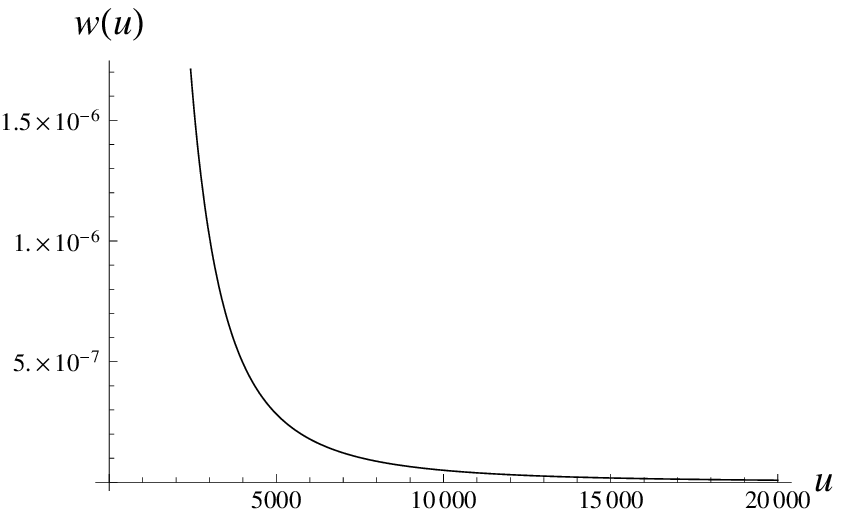}
\includegraphics[scale=.8]{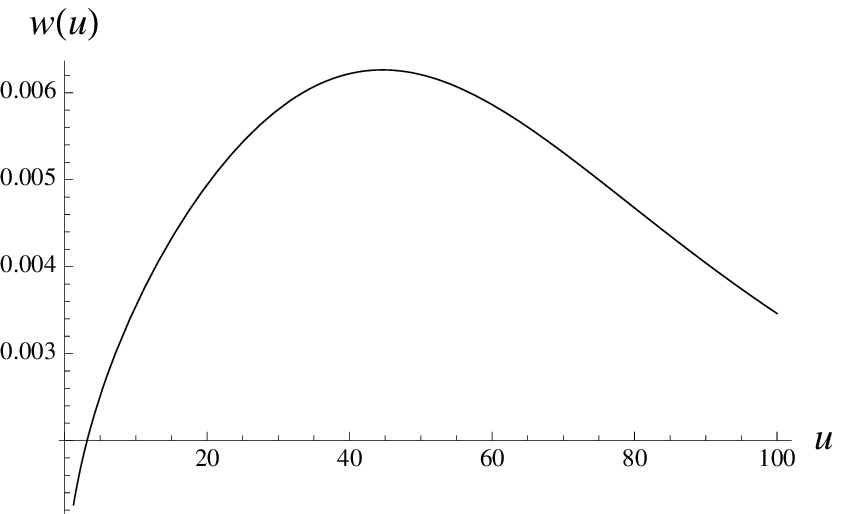}
\caption{Plot of the charged W boson - numerical. 
The second plot expands the small $u$ region.}
\end{center}
\end{figure}

\begin{figure}[htbp]
\begin{center}
\includegraphics[scale=.8]{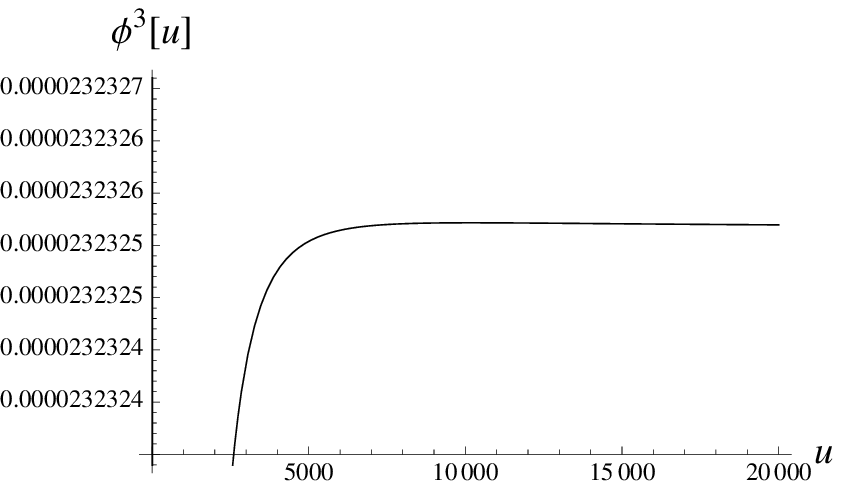}
\includegraphics[scale=.8]{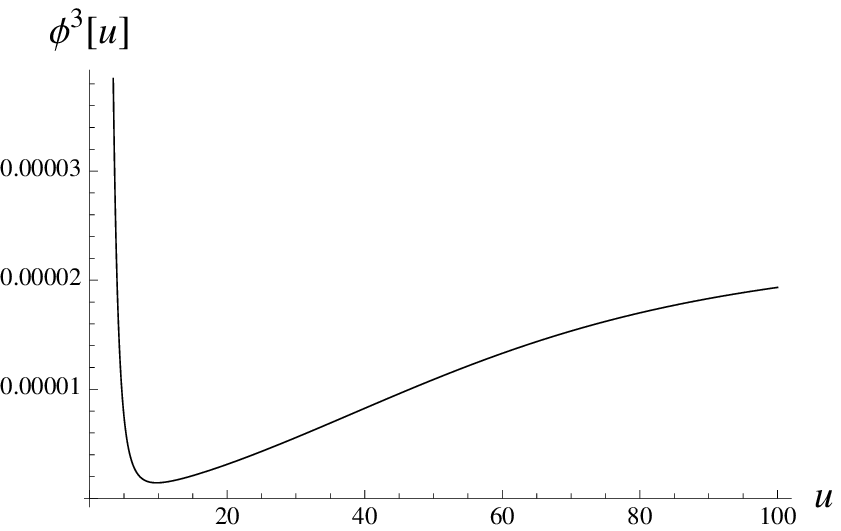}
\caption{Plot of $\phi ^3$ - numerical.
The second plot expands the small $u$ region.}
\label{t3}
\end{center}
\end{figure}

\begin{figure}[htbp]
\begin{center}
\includegraphics[scale=.8]{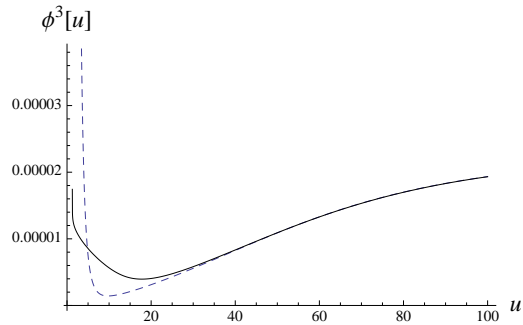}
\caption{Comparison of analytical (dashed) and numerical
solutions for $u<100$. The agreement is good for large $u$ as
expected. }
\end{center}
\end{figure}

\begin{figure}[htbp]
\begin{center}
\includegraphics[scale=.8]{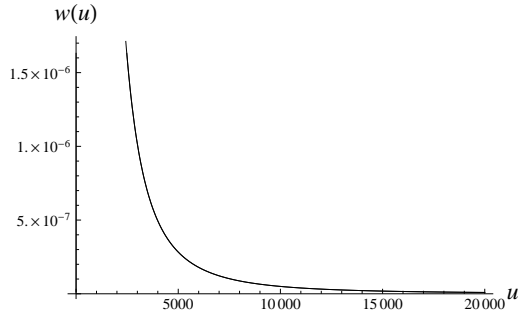}
\caption{Comparison of analytical (dashed) and numerical
solutions for asymptotic values of $u$. The dashed line is not visible as the agreement is good. }
\end{center}
\end{figure}

A technical comment about the numerical solutions is that
equation (\ref{w-}) is not automatically implemented. However
gauge invariance of the action implies the following relation
between the equations: $D_\mu {\partial \cal L \over \partial
A_\mu} +i e [\phi,{\partial \cal L \over \partial \phi}] = 0
\;$.  Thus, in this context it means that
\be
\du ({\partial S\over \partial A_u^-})= 2i A_u^+ ({\partial
S\over \partial A_u^3})+i A_0^3({\partial S\over \partial
A_0^-}) + i (\phi _3 ({\partial S\over \partial \phi^-})-2\phi^+
({\partial S\over \partial \phi^3}))
\ee
This means that if the other equations are satisfied (\ref{w-})
must evaluate to a constant. Thus for some choice of boundary
conditions this constant will become zero and the equation will be satisfied. 
Thus we have scanned over different values of the boundary values 
till (\ref{w-}) evaluates to zero. The result of this is an exact solution that
satisfies all the equations of motion. This is shown in Figures 5,6,7 and 8.
The value of the $SU(2)$ fields for this solution are very small in the entire region
and hence the Yang-Mills approximation to DBI is reliable.
This then establishes unambiguously that a condensate is indeed formed.

\section{Phases as a function of $\mu$ and $T$}

This section summarizes the phase structure (at zero temperature) based
on the calculations of the earlier sections.
To study the thermodynamics we go to Euclidean section by
setting $it = t_E$ where $t_E$ has a period $\beta ={1\over T}
\;$. To compare the phases at zero temperature we consider the
Euclidean action with $\beta = \infty$, which effectively means
consider the coefficient of $\beta$.  We have a background field
$A_0 = A_t$. Wick rotation converts this to $A_{t_E}$ which
satisfies $A_{t_E} d t_E = A_t dt$. Thus $A_{t_E} = {A_t \over
i} \;$. When one does the Euclidean functional integral we
analytically continue to real $A_{t_E}$ and integrate over real
$A_{t_E}$ field configurations. However if there is a background
{\em real} $A_t$ (that is a solution of the classical Minkowski
equations with a {\em real} charge), this has to be put in as an
{\em imaginary} background value for $A_{t_E} \;$.  Thus if one
wants to perform the functional integral at one loop, then one
has to work with real $A_{t_E}$ fluctuating about an imaginary
background value. However at the classical level, we need not do
all this - we can just plug in the Minkowski solution into the
action without introducing $A_{t_E} \;$.

Since the configurations depend only on $u$, we need only worry
about the $u$ integral. Thus we need to look at $S[A^3_0(u), \;
\tau (u)]$ evaluated on the solutions.

As a first step we just consider the solutions of the
Dirac-Born-Infeld equations considered in Section 1 where the sources
are localized as $\delta (u-u_c)$. This can tell us whether the
phase with $t_3$ number being non zero is preferred over the
phase with net $t_3$ number being zero.  We remind the reader
that we are referring to the $t_3$ component of the $U(1) \times
SU(2) \;$. ($t_a \;$ are the $SU(2)$ generators.) We assume that
the $U(1)_B$ baryon charge is already non zero and we have the
cusp configuration. When $t_3$ number is non zero, one can
further ask whether there is a $t_3$ number violating
condensate. This has to be asked within the Yang-Mills
approximation. This is the second step.

The Dirac-Born Infeld action with source is (see equation
(\ref{SSDBI}))
\be 
S_{DBI}=-\tilde T \int ~ d \sigma~ 
{u^{5 \over 2} \over \sqrt{f}} \; \; [ u'^2 + u^3 f^2 \tau'^2 
- f (\partial_\sigma a_0)^2 ]^{1 \over 2} 
- n_B \tilde T a_0(u_c)
\ee
and the solution for $a_0$ is
\be
(\partial_u a_0)^2 = {d_0^2 \; (1 + u^3 f^2 {\tau '}^2 )
\over f \; (u^5 + d_0^2)}
\ee

This determines $a_0(u)$ in terms of $d_0$. In our case we have
two branes. We write $d_1, d_2$ corresponding to the two branes
as: 
\[
d_{1} = d_0 + {d_3\over 2} ~~~~
d_{2} = d_0 - {d_3\over 2} 
\]
where $d_3$ is the $t_3$ charge that we are concerned with. The
chemical potential $\mu_3 = A^3_0 (\infty)$. We will fix a gauge
by setting $A^3_0 (u_c) = 0 \;$.\footnote{ Normally when the
gauge field extends all the way to the horizon, one usually sets
the gauge field to zero at the horizon. The gauge field here
lives on the D8 brane, which does not extend beyond $u = u_c
\;$.}
 
We have two branes with different profiles $\tau (u)$, which in
turn depend on the constants $d$ and $c$, see equation
(\ref{tau'}) . We need to impose that $u_c$ is the same for both
branes. So if $d_1,c_1$ are given, then $u_c$ is fixed and so
$c_2$ is fixed once $d_2$ is. Thus we Legendre transform $\tau
'(u)$ in terms of $c_0$. We can also substitute the solution for
$a_0(u)$ and $\tau(u)$. Thus we find that the Legendre
transformed (with respect to $\tau$) is 
\[
\Omega (\mu, \; c_0) = {f u^8 - c_0^2 \over f u^{3 \over 2}
\; \sqrt{D (d_0, c_0)}} \; \; \; , \; \; \; \; 
D (d_0, c_0) = u^8 f + d_0^2 u^3 f - c_0^2 \; \; .
\]
Generalizing to two branes one finds:
\[
\Omega (\mu_0, \mu_3, c_0) = \int du \; \left( 
{f u^8 - c_1^2 \over f u^{3 \over 2} \; \sqrt{D (d_1, c_1)}}
+ {f u^8 - c_2^2 \over f u^{3 \over 2} \; \sqrt{D (d_2, c_2)}}
\right) 
\]
where $d_{1} = d_0 + {d_3\over 2}$ and $d_{2} = d_0 - {d_3\over
2} \;$.

One can compare this action with the action with $d_3 = 0$ and
the same value of $\mu_3$ at the boundary. If $d_3 = 0$ then the
branes are right on top of each other and there is no
separation. Thus not only is there no $t_3$ charge, there is
also no $t_3$ number violating condensate. We have an unbroken
$SU(2)$ gauge symmetry. If $d_3 = 0$ then $A^3_0$ is constant
everywhere and is pure gauge. The action is
\[
\Omega_0 (\mu_0, \mu_3, c_0) = 2 \; \int du \; 
{f u^8 - c_0^2 \over f u^{3 \over 2} \; \sqrt{D (d_0, c_0)}}
\; \; . 
\]

\begin{figure}
\begin{center}
\includegraphics[scale=.8]{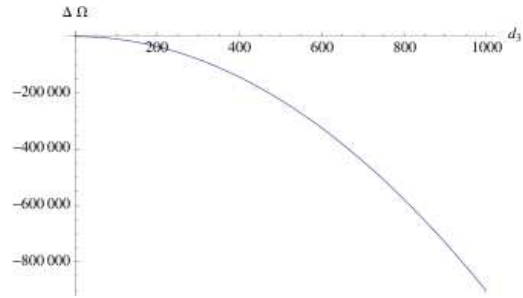}
\caption{Free Energy difference: $c_{1}=10^3,d_{1}=10^3$
\label{fd3}}
\end{center}
\end{figure}
\begin{figure}
\begin{center}
\includegraphics[scale=.8]{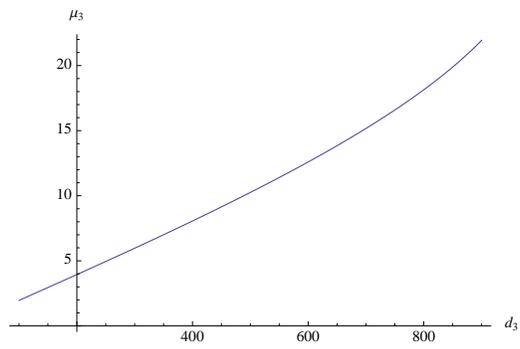}
\caption{$\mu_3$ vs $d_{3}$ with $c_{1}=10^3,d_{1}=10^3$
\label{mu3d3}}
\end{center}
\end{figure}

The precise relation between $\mu$ and $d_0$ can be obtained by
first calculating the free energy $F(d_0, c_0)$ with fixed $d_0$
(Legendre transform of $\Omega$ with respect to $\mu \;$, or
equivalently, of $S$ with respect to $a_0 \;$). 
\[ 
F(d_0, c_0) = \tilde T \; \int_{u_c}^\infty du \; 
{\sqrt{D(d_0, c_0)} \over f(u) u^{3 \over 2}} \; \; 
+ \; {\tilde T \; d_0 \over 3} \; u_c \; \; . 
\] 
We have included the  D4-brane action of the $U(1)$ charges. This gives
\[ 
\mu = {\partial F \over \partial d_0} = \tilde T
\int_{u_c}^\infty du ~ \du a_0 \; + \; {\tilde T \; u_c \over 3}
= \tilde T \int _{u_c}^\infty du ~ {d_0 u^{3 \over 2} \over
\sqrt{D(d_0, c_0)}} \; + \; {\tilde T \; u_c \over 3} \; \; .
\] 
The free energy difference is $\Omega (d_1) + \Omega(d_2) - 2 \;
\Omega (d_0) \;$. This has to be plotted for different $\mu_3$
keeping fixed $\mu_0=\mu_1+\mu_2$.  It is simpler to plot the
free energy $\Omega$ as a function of $d_3$ and give a separate
plot giving $\mu_3$ versus $d_3$.

The chemical potential is given by:
\[ 
\mu_3 = \mu _1 -\mu_2 = \int _{u_c}^\infty du ~ \left( 
{d_1 u^{3 \over 2} \over \sqrt{D(d_1, c_1)}} \; 
- \; {d_2 u^{3 \over 2} \over \sqrt{D(d_2, c_2)}} \right) 
\]

Figure {\bf \ref{mu3d3}} shows a plot of $\mu_3$ vs $d_3$ for
the separated D8 brane solution (i.e. with a $t_3$ number) for a
sample value of $c_0, d_0 \;$. The free energies are plotted as
a function of $d_3$ in Figure {\bf \ref{fd3}}. We see that for
all values of $d_3$ the free energy difference is negative. For
e.g. for $d_3=10$ (so that $d_3<<d_0$), we find that $\Delta
\Omega \simeq -89.97 \;$. Thus one infers that as $\mu_3$
increases so does $d_3$ (see Figure {\bf \ref{mu3d3}}), and so
(see Figure {\bf \ref{fd3}}) one has a phase transition to the
phase with non zero $t_3$ number.

We have thus established that when there is a non zero chemical
potential for $t_3$ the phase with a finite number density of
$t_3$ charge is favored. We also have exact (numerical)
solutions for this phase which have charged fields condensing. A
calculation of the free energy difference for the given exact
solution gives $\simeq - 4.40 * 10^{- 4}$ and shows that this
condensate is favored over the trivial solution (with condensate
being zero). Thus putting these results together we have
established that a BCS condensate is formed and we have a
holographic description of the ground state.  This is the main
conclusion of this paper.  We now discuss below the implications
of this for the boundary theory.

Our experience with the boundary BCS theory says that there
should be a condensate, since when there is a finite number
density of fermions, and hence a Fermi Surface, one expects the
BCS instability. This should thus reflect itself in a $t_3$
number violating condensate of charged fields.  As mentioned
above the existence of this condensate is proved by the
holographic calculation.

The nature of the boundary condensate also needs to be
determined. As discussed in the Introduction the most plausible
candidate seems to be a BCS like Cooper pair $BB$ if $N_c$ is
odd. We are at zero temperature and in the confining phase in
the boundary. $B = \epsilon_{a_1, a_2, \cdots, a_{N_c}} u^{a_1}
d^{a_2} u^{a_3} d^{a_4} \cdots u^{a_{N_c}}$. If $N_c$ is even,
one could have scalar baryons with $t_3 = + 1$ condensate. This
would be analogous to a BEC because the particles are strongly
bound already, before they condense.

The effect of this condensate should be visible in the
conductivity. Thus if one calculates the current current
correlator one should find a gap in the imaginary part
(equivalently in the real part of the conductivity). This is a
problem for the future. In the next section we take a first step
in this direction by considering the conductivity of the
Sakai-Sugimoto model without any $SU(2)$ breaking.


\section{Conductivity}

Consider the DBI action given by equation (\ref{SSDBI}) with
$n_B = 0$ and $\sigma = u \;$,
\be
S_{DBI}=-\tilde T \int d^4 x ~ d u ~ 
{u^{5 \over 2} \over \sqrt{f}} \; \; 
[1 + u^3 f^2 \tau'^2 - f (\partial_u a_0)^2 ]^{1 \over 2}
\; \; . 
\ee
The equations of motion are
\beqa
\frac {u^{11} f^3 \; \tau'^2} 
{1 + u^3 f^2 \tau'^2 - f (\partial_u a_0)^2} & = & c_0^2
\nonumber \\
\frac {u^5 f \; (\partial_u a_0)^2} 
{1 + u^3 f^2 \tau'^2 - f (\partial_u a_0)^2} & = & d_0^2
\eeqa
where $c_0$ and $d_0$ are constants. These equations imply, 
see equations (\ref{tau'}), 
\be \label{psol}
(\partial_u a_0)^2= {d_0^2 \; u^3 \over D} 
\; \; \; , \; \; \; \; 
(\partial_u \tau)^2 = {c_0^2 \over u^3 f^2 D}
\ee
where $D = u^8 f + d_0^2 u^3 f - c_0^2 \;$ and, hence, that 
\begin{equation}\label{L0}
{\cal L} =  {u^{5 \over 2} \over \sqrt{f}} \; \; 
[1 + u^3 f^2 \tau'^2 - f (\partial_u a_0)^2 ]^{1 \over 2}
= \frac{u^{13 \over 2}}{\sqrt{D}} \; \; . 
\end{equation}

The $D8$ brane ends at $u = u_0$ where $\pa_u \t \to \infty$
{\em i. e.} $D(u_0) = 0 \;$. This gives $u_0$ in terms of other
parameters as
\be
(u_0^5 + d_0^2) \; u_0^3 f_0 = c_0^2 
\; \; , \; \; \; f_0 = f(u_0) \; \; . 
\ee


\subsection{Fluctuation about the background}

Let us turn on the $U(1)$ gauge field $a_2(\sigma^\mu) $ along
the $\sigma^2 = x \;$ direction of the $D8$ brane worldvolume.
We will consider this as small fluctuation around the background
described in previous section and neglect any back reaction. For
simplicity let us first assume $a_2$ to be a function of $t$ and
$u$ only. We will also assume $a_2 (t, u) = a(u) e^{-i \o t}$.
Let a dot $(\dot{ })$ and a prime ($'$) denote $t$ and $u$
derivatives.

For matrices $M$ and $\delta M$, where $\delta M$ is assumed to
be small, we have
\beqa
\frac {\sqrt{det \; (M + \delta M)}} {\sqrt{det \; M}} & = & 
1 + \frac{1}{2} \; tr (M^{- 1} \delta M)
+ \frac{1}{8} \; \left(tr (M^{- 1} \delta M) \right)^2 
\nonumber \\
& & - \frac{1}{4} \; tr (M^{- 1} \delta M)^2 
+ {\cal O}(\delta M)^3 \; \; .
\eeqa
Applying this relation to the Lagrangian ${\cal L}_{DBI} =
\sqrt{det \; (G + F + \delta F )} \;$ gives
\beqa
{\cal L}_{DBI} \approx \mathcal{L} - \f{1}{4}
\mathcal{L} ~~ tr((G + F)^{-1} \d F)^2
+ {\cal O}(\delta F)^3 
\eeqa 
where ${\cal L} = \sqrt{det \; (G + F)} \;$. For the problem
considered here, ${\cal L}$ is given explicitly in equation
(\ref{L0}), and we have $\; tr ((G + F)^{-1} \d F) = 0 \;$. We
can construct $3 \times 3$ matrices out of $(G + F)$ and $\d F$
to compute the traces, as $\d F$ is of the form,
\be
\d F = \left(
  \begin{array}{cccc}
    0 & 0 & \dot a_2 & \\
     0 & 0 & a_2' & \mbox{\Huge $0$}\\
     -\dot a_2 & - a_2'& 0 & \\
 & \mbox{\Huge $0$} & &  \\
& & & \mbox{\Huge $0$} 
  \end{array}
\right) \; \; . 
\ee
Let us call the truncated blocks of $(G + F)$ and $\d F \;$ as
$B$ and $\d B$ respectively,
\beqa
B &=& \left(
  \begin{array}{ccc}
   -u^{3 \over 2} & -a_0' & 0  \\
   a_0' & \f{1}{u^{3 \over 2} f} (1 + u^3 f^2 \t'(u)^2)  & 0 \\
  0 & 0& u^{3 \over 2} 
  \end{array}
\right) \non \\
\d B &=& \left(
  \begin{array}{ccc}
   0 & 0 & \dot a_2  \\
   0 & 0 & a_2' \\
  -\dot a_2 & -a_2'& 0 
  \end{array}
\right)
\eeqa
Then,
\be
tr((G + F)^{-1} \d F)^2 = \f{1}{\mathcal{L}^2} 
\; \f{2 u^2}{f} \;
((1 + u^3 f^2 \t'^2) (\dot{a}_2)^2 - u^3 f (a'_2)^2)
\ee
Then the Lagrangian $\mathcal{L}_f$ for the fluctuations is
given by
\be\label{71}
\mathcal{L}_f = \f{1}{2} \lb \f{(u^5 + d_0^2)}{\sqrt{u^3 D}}
\; (\dot{a}_2)^2 - \sqrt{D \over u^3} \; (a'_2)^2 \rb \; \; . 
\ee
The equation of motion is then given by
\be
\left( \sqrt{ D \over u^3} \; a'_2 \right)'  
- \f{(u^5 + d_0^2)}{\sqrt{u^3 D}} \; \ddot{a}_2 = 0 \; \; . 
\ee
Setting $a_2 (t, u) = a(u) e^{-i \o t}$, we get 
\be \label{diffeq1}
\left( \sqrt{ D \over u^3} \; a' \right)'  
+ \f{(u^5 + d_0^2)}{\sqrt{u^3 D}} \; \omega^2 = 0 \; \; . 
\ee

The equation of motion is same for the world volume fluctuations
considered above for both $D8$ and $\overline{D8}$ branes.  In
the absence of charges, continuity of flux would require that
the fields continue in the same direction along the other brane
and reemerge on the boundary. So we can consider $a(u)$ to be
the field on both $D8$ and $\overline{D8}$ branes, but $u$ is a
bad co-ordinate choice for such a representation. We can
consider a change of co-ordinate given by $y^2 = 1 -
\f{u_0}{u}$.  $y \in (0,1)$ corresponds to brane and $y \in (0,-
1)$ anti-brane. The brane and anti-brane are joined at $y = 0
\;$, the $y = \pm 1$ are corresponds to the intersection of the
brane and anti brane with $D4 \;$. We will call $y = 1$ boundary
and $y = - 1$ the ``horizon''.  The differential equation will
be solved with ``in-going boundary condition at the horizon'',
and we will use ``AdS/CFT correspondence'' to evaluate the
boundary Green's function.

This choice of boundary condition does not have a rigorous
justification. However we can argue heuristically. We want
boundary conditions corresponding to the normal modes or more
correctly quasi normal modes. The eigenvalues then give the
poles of the Green function. We motivate boundary conditions as
follows. In the case of the black hole, in the AdS/CFT context,
these boundary conditions were first introduced in \cite{KS1,HH}
where it was argued that at the horizon, ingoing boundary
conditions are useful because with the usual Dirichlet/Neumann
boundary condition the eigenvalues are strictly real. The latter
would give real poles for the boundary Green function and would
not describe thermalization which requires an imaginary part for
the poles. Ingoing boundary condition is one option that does
give an imaginary part to the pole location.  Furthermore
ingoing boundary condition has the reasonable physical
interpretation of complete absorption by the black hole.

In the present case also we only have a few reasonable boundary
conditions and we pick one that seems to give the right pole
structure. Ingoing, is one such and corresponds to perfect
absorption. Are there situations other than a black hole horizon
where one expects complete absorption? The answer is yes.
Consider electromagnetic waves in a wave guide with an
oscillator source at one end and the other end open and
radiating via an antenna into three dimensional space. The cross
section of the wave guide is two dimensional. It is well known
that when there is impedance matching the entire output of the
wave guide is radiated and there is no reflection at the
boundary. The radiating antenna behaves like a perfect absorber
at the end of the wave guide. The correct boundary condition
inside the wave guide in such a situation is the ingoing
one. Thus perfect absorption is not unknown outside the black
hole context. In our case we have a lower dimensional wave
propagation channel opening out into a higher dimensional space
- the cross section of the D8 brane is effectively three
dimensional and is connecting to a four dimensional D4 brane.
This gives a possible physical motivation for considering
ingoing boundary conditions in the present situation.

We will set $u_{KK} = 1$ in all the future calculations. The
differential equation for $a(y)$ near ``horizon'', {\em i.e.}
near $y = - 1 \;$, is given by
\be
(1 + y) \; \frac{d^2 a}{d y^2}
- \f{1}{2} \; \frac{d a}{d y} + \f{2 \o^2}{u_0} a = 0
\ee
which gives the near-horizon behavior of $a(y)$ as,
\be
a \sim e^{\pm  i \f{2 \sqrt{2} \o}{\sqrt{u_0}}\sqrt{1+y}}
\ee
The solution $a \sim e^{- i \f{2 \sqrt{2} \o} {\sqrt{u_0}}
\sqrt{1 + y}}$ gives the in-going wave. Let $a(y) = e^{- i \f{ 2
\sqrt{2} \o} {\sqrt{u_0}} \sqrt{1 + y}} \rho(y) \;$. Then the
differential equation for $\rho(y)$ is solved numerically to
obtain $a(y)$, with boundary condition that $\r(y_h) = 1 + \f{2
\sqrt{2} i \o}{\sqrt{u_0}} \sqrt{1 + y_h}$ and $\r'(y_h) =
\f{\sqrt{2}~i~\o}{\sqrt{u_0} \sqrt{1 + y_h}} \;$, where $y_h$ is
a cut-off near $y = - 1 \;$. The expression for $\r(y_h)$ is
obtained by doing a series solution of the differential equation
near $y = - 1 \;$.

The differential equation for $a(y)$ near ``boundary'', {\em
i.e.}  near $y = + 1 \;$, is given by
\be
(1 - y) \; \frac{d^2 a}{d y^2}
+ \f{1}{2} \; \frac{d a}{d y} + \f{2 \o^2}{u_0} a = 0
\ee
The solution for $a(y)$ near boundary, {\em i.e.}  near $y = + 1
\;$, behaves as
\begin{eqnarray}
a(y) & \sim & A \; 
(1 + A_1 (1 - y) + A_2 (1 - y)^2 + \cdots) \nonumber \\
& & + B \; (1-y)^{\f{3}{2}} \; 
(1 + B_1 (1 - y) + B_2 (1 -y)^2 +\cdots) 
\end{eqnarray}
where $A$ and $B$ are arbitrary constants determined by boundary
conditions. $A_i$ and $B_i$ are constants determined in terms of
parameters of the differential equation.  So, the Green's
function is given by,
\be
G_R (\o, u_0, d_0) \sim \f{B}{A} = \lim_{y \to 1} \; 
\f{\f{4}{3} ~ \sqrt{(1-y)}}{a(y)} \; \frac{d^2 a}{d y^2} \; \; . 
\ee
The conductivity can then be obtained by
\be
\s = \f{G_R}{i~\o} \; \; . 
\ee

\begin{figure}[h]
\begin{center}
\includegraphics{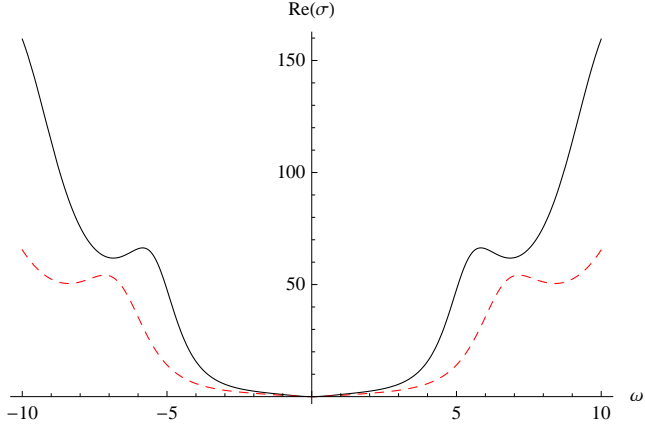}
\caption{Real part of Conductivity with $d_0=0$: $u_0=10$
(Black, continuous) and $u_0=15$ (Red, dashed)
\label{ReCond0m0d}}
\end{center}
\end{figure}

\begin{figure}[h]
\begin{center}
\includegraphics{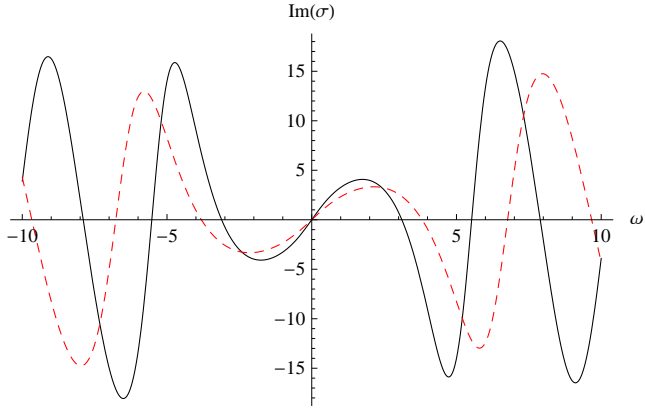}
\caption{Imaginary part of Conductivity with $d_0=0$: $u_0=10$
(Black, continuous) and $u_0=15$ (Red, dashed). 
\label{ImCond0m0d}}
\end{center}
\end{figure}

Figures {\bf \ref{ReCond0m0d}} and {\bf \ref{ImCond0m0d}} show
variation of conductivity with frequency. We have set $d_0 = 0$
in our conductivity calculations as the variation of conductivity with
$d_0$ is very small.

\subsection{Modified Sakai-Sugimoto model}

\begin{figure}[h]
\begin{center}
\includegraphics[scale=.8]{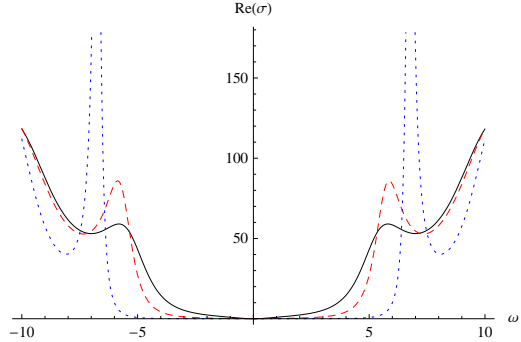}
\caption{Real part of Conductivity with $d_0=0$, $u_0=10$: $m_0=0$ (Black,
continuous), $m_0=5$ (Red, dashed) and $m_0=10$ (Blue, dotted). As we increase the mass $m_0$,
the peak height increases. The peak for $m_0 = 10$ is outside the range of the ordinate
value shown in the figure.
\label{ReCondm0d}}
\end{center}
\end{figure}

\begin{figure}[h]
\begin{center}
\includegraphics{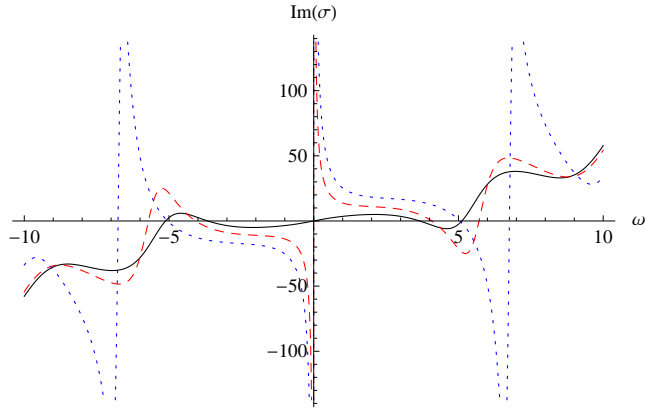}
\caption{Imaginary part of Conductivity with $d_0=0$, $u_0=10$: $m_0=0$
(Black, continuous), $m_0=5$ (Red, dashed) and $m_0=10$ (Blue,
dotted). The peak for $m_0 = 10$ is outside the range of the ordinate
value shown in the figure\label{ImCondm0d}}
\end{center}
\end{figure}

In the original paper, the scalar bi-fundamental tachyon field
arising from open string between $D8$ and $\overline{D8}$-brane
was neglected. Later various authors~\cite{DN, BLL} have
included the contribution of the tachyon field which is
expected to be the origin of quark mass and condensate. The
effect of tachyon field for the case of gauge field fluctuation
is that the fluctuations become massive, with mass proportional
to the tachyon field. The tachyon field has an approximate
behavior like $e^{-\f{u}{u_0}}$ asymptotically. We will assume
the mass term corresponding to the $a_2$ fluctuation is
proportional to $m_0^2 e^{-\f{u}{u_0}} \;$, to get a qualitative
nature of the effect of tachyon on the conductivity. More
rigorous analysis needs to be done for quantitative results. Let
us consider the modification of equation (\ref{diffeq1}) given
by
\be 
\left( \sqrt{ D \over u^3} \; a' \right)'  
+ \f{(u^5 + d_0^2)}{\sqrt{u^3 D}} \; 
(\omega^2 - m_0^2 e^{-\f{u}{u_0}}) = 0 \; \; . 
\ee
A similar analysis, as in the previous section, can be done to
obtain conductivity. Figures {\bf \ref{ReCondm0d}} and {\bf
\ref{ImCondm0d}} show the conductivity for various values of
$m_0 \;$. We see that with a non-zero value of $m_0$ the
conductivity develops a mass-gap. Effect of $d_0$ on the
conductivity is very small, and we have set $d_0 = 0$ for all
our numerics. As we increase $m_0$, both mass gap and peak heights
increases. Also, for any finite value of $m_0$, a pole appears at $\o=0$
for the imaginary part of conductivity (fig. {\bf \ref{ImCondm0d}}), which implies a 
delta function peak for real part of the conductivity by Kramers-Kronig relation.
The delta function function peak can not be captured by numerics.

\section{Summary and Conclusions}

In this paper we have attempted to find a holographic
description of the BCS Cooper pairing instability and thence
describe the strong coupling microscopic BCS phenomenon. The
analogy with chiral symmetry breaking in QCD was exploited and
the Sakai Sugimoto model with finite number density of flavored
fermions was studied. The background charge corresponds to a
$U(1)_3$ embedded in the $SU(2)_{flavour}$. The flavor $SU(2)$
was broken to $U(1)_3$ by the background and it was shown that in
this situation there is a tachyonic instability that causes
fields charged under this $U(1)_3$ to condense. This is the bulk
description of the BCS instability.  Analytic solutions
describing this condensate were given in flat space.  The
stability properties were also studied. The curved space
equations were then solved analytically in some approximation.
The exact solutions have also been done numerically.  The free energy difference shows that
this solution is favored over the trivial solution. This solution then describes the
BCS ground state of the superconductor.

Once we have this solution one should be able to calculate
various quantities such as the AC conductivity using the
AdS/CFT dictionary. However there are two complications. One is
that the bulk is not asymptotically AdS.  The second is that the
flavor branes do not extend all the way to the interior. Thus
the usual ``infalling" boundary conditions prescription has to
be modified. A preliminary exploration of this problem in the
original Sakai-Sugimoto model was done in Section 7. The chiral
current Greens function was calculated. It has the expected
properties of a mass gap.

The detailed analysis of this in the context of the
Sakai-Sugimoto model as well as in the modified form in this
paper need to be studied. Another important question is to look
for the bulk signature of the Fermi surface in the boundary.

We hope to return to these questions soon.

{\bf Acknowledgements:} We thank G. Baskaran and R. Shankar for
many useful discussions. We also thank the unknown referee of Nuclear Physics B
for very useful comments.

\newpage


\appendix

\section{$D8$ brane with non zero electric field}\label{appkr1} 


In this Appendix, we formulate the equations describing the
profile of $D8$ brane in the background of $D4$ branes, and with
a non zero world volume electric field turned on. 

We write the Dirac-Born-Infeld part of the $D8$ brane action in the
ten dimensional background of $D4$ branes as
\[
S_{DBI} = - C_9 \; \int d^9 \sigma \; g_s e^{- \Phi}
\; \sqrt{- det \; G} 
\]
where the components $G_{\mu \nu}$ of the matrix $G$ are given
by
\[
G_{\mu \nu} = g_{M N} \; \partial_\mu X^M \partial_\nu X^N 
+ F_{\mu \nu} 
\]
with $g_{M N}$ being the ten dimensional background metric, 
\[
X^M = (x^0, u, x^1, x^2, x^3, \tau, \theta^1, \cdots, \theta^4)
\]
are the ten dimensional coordinates with $\tau \sim \tau + 2 \pi
r_4$, and $F_{\mu \nu} = \partial_\mu a_\nu - \partial_\nu a_\mu
$ is the $U(1)$ gauge field strength on the brane. We choose the
nine dimensional brane coordinates $\sigma^\mu$ to be given by
\[
\sigma^\mu = (x^0, \sigma, x^1, x^2, x^3, \theta^1, \cdots,
\theta^4) \; \; . 
\]

We consider the case where $a_0$ is the only non vanishing
component of the $U(1)$ gauge field and $(u, \tau, a_0)$ depend
on $\sigma$ only. The functions $u(\sigma)$ and $\tau(\sigma)$
describe the $D8$ brane profile in the $(u, \tau)$ plane. It
then follows that
\[
G_{\sigma \sigma} = g_{u u} \; u_\sigma^2 + g_{\tau \tau} \;
\tau_\sigma^2 
\; \; \; , \; \; \;
G_{0 \sigma} = - G_{\sigma 0} = F_{0 \sigma} = - (a_0)_\sigma
\; \; ,
\]
and $G_{\mu \nu} = g_{\mu \nu}$ otherwise. The subscripts
$\sigma$ on $(u, \tau, a_0)$ denote their \\ 
$\sigma-$derivatives.

In order to solve the equations of motion which follow from the
action $S_{DBI}$ and to obtain the $\sigma-$dependence of $(u,
\tau, a_0)$, note that $S_{DBI}$ now resembles a world line
action
\[
S_{DBI} \sim \; \int d \sigma \; \sqrt{h_{\alpha \beta} \;
\xi^\alpha_\sigma \xi^\beta_\sigma }
\]
where $\xi^\alpha = (u, \tau, a_0)$, the subscript $\sigma$ on
$\xi$ denotes its $\sigma-$derivative. For our case, $g_{M N}$
is diagonal and we have $h_{\alpha \beta} \equiv diag \; (h_1,
h_2, h_3) \;$ where
\begin{eqnarray*}
h_1 & = & g_s^2 e^{- 2 \Phi} \; 
(- \; det \; g_{\mu \nu}) \; \vert_{\sigma = u} \\
h_2 & = & g^{u u} g_{\tau \tau} \; h_1 \\
h_3 & = & g^{u u} g^{0 0} \; h_1 
\end{eqnarray*}
and depend on $\xi^1 = u$ only. The consequent `geodesic'
equations then give
\begin{eqnarray}
& & h_2 \; \tau_\sigma = c_0 \label{tausigma} \\
& & h_3 \; (a_0)_\sigma = d_0 \label{a0sigma} \\
& & h_1 \; u_\sigma^2 + h_2 \; \tau_\sigma^2 
+ h_3 \; (a_0)_\sigma^2 = E \label{usigma} \\
& & 2 \; h_1 \; u_{\sigma \sigma} + (h_1)_u \; u_\sigma^2 
- (h_2)_u \; \tau_\sigma^2 
- (h_3)_u \; (a_0)_\sigma^2 = 0 \label{usigmasigma} 
\end{eqnarray}
where $c_0$, $d_0$, and $E$ are constants and the subscripts
$u$ on the $h$s denote their $u-$derivatives. It follows from
the above equations that
\begin{equation}\label{usigmau}
u_\sigma^2 = \frac{1}{h_1} \; \left(E - \frac{c_0^2}{h_2} 
- \frac{d_0^2}{h_3} \right) \; \; .
\end{equation}

The right hand side of the above equation is a function of $u
\;$. Generically the location $u = u_0$ of its zero, if exists,
denotes a turning point for $u(\sigma)$. Thus if $u(\sigma) \to
u_0$ from above as $\sigma \to \sigma_0$ then, generically, $u$
starts increasing as $\sigma$ increases beyond $\sigma_0 \;$
which can be seen from $u_{\sigma \sigma}$ equation. Also,
generically, the evolution of $\tau (\sigma)$ and $a_0 (\sigma)$
is monotonous across $\sigma_0 \;$. In the context of $D8$
branes in the background of $D4$ branes, $u_0$ will denote the
lowest point of $D8$ brane profile in the $(u, \tau)$
plane. Also, $(a_0)_\sigma$ equation above with $d_0 \ne 0$
shows that such a $D8$ brane can support a non trivial $U(1)$
gauge field on its worldvolume. It is now straightforward to
obtain the evolution in terms of $u$ alone using $\tau_u =
\frac{\tau_\sigma}{u_\sigma}$ and similarly for $(a_0)_u \;$.
Using $F_{0 u} = - (a_0)_u \;$ and the expressions for $h_2$ and
$h_3 \;$ in equation (\ref{usigma}), we have
\begin{equation}\label{Delta}
\Delta \equiv 1 + g^{u u} g_{\tau \tau} \; \tau_u^2 
+ g^{0 0} g^{u u} \; F_{0 u}^2 = \frac{E}{h_1 \; u_\sigma^2}
\end{equation}
where $u_\sigma^2$ is given in equation (\ref{usigmau}).

Writing the ten dimensional background fields of $D4$ branes as
\begin{eqnarray}
d s_{10}^2 & = & g_{M N} d x^M d x^N  \nonumber \\ 
& = & H^{- \frac{1}{2}} \left( - (d x^0)^2 
+ \sum_{i = 1}^3 (d x^i)^2 + f d \tau^2 \right) 
+ H^{\frac{1}{2}} \left( \frac{d u^2}{f} 
+ u^2 d \Omega_4^2 \right) \nonumber \\ 
\label{d4metric} 
\end{eqnarray}
and $e^\Phi = g_s \; H^{- \frac{1}{4}} \;$ where $H = \frac {1}
{u^3} \;$, and $f = 1 - \frac{u_{K K}^3}{u^3} \;$, we have
\[
(h_1, h_2, h_3) = u^8 \; 
\left( \frac{H}{f}, \; f, \; - H \right) \; \; , \; \; \;
u_\sigma^2 = \frac{D}{H u^{16}} \; \; , \; \; \;
\Delta = \frac{u^8 f \; E}{D} 
\]
where $D = (E u^8 + \frac{d_0^2}{H}) f - c_0^2 \;$, and
\[
(a_0)_u^2 = \frac{d_0^2}{H \; D} 
\; \; , \; \; \; \; 
\tau_u^2 = \frac{c_0^2 \; H}{f^2 \; D} \; \; . 
\]
The expressions for $(a_0)_u^2 \;$ and $\tau_u^2$ given in
equations (\ref{tau'}) correspond to the choice $E = 1 \;$.


\section{Embedding of the $SU(2)$ group in $SU(2N_d)$}

We have two sets of coincident branes. For concreteness and clarity of presentation let us assume that each
coincident set has three branes, i.e $N_d=3$. (The minimum required is two). So we have six branes in all. Our configuration breaks $U(1) \times SU(6)$ down to
$U(1)\times U(1) \times SU(3)\times SU(3)$. We have 36 generators in $U(6)$. We can write them using the direct product notation. Let $\l$ stand for the Gell-Mann Lambda matrices for the $SU(3)$ part: $\l ^{1,2,..8}$ and $\l ^0$ be the identity matrix.\[ \l^0=\left( \begin{array}{ccc} 1&0&0\\0&1&0\\0&0&1 \end{array}\right)\]

 Similarly $\s ^{1,2,3}$ are the $SU(2)$ matrices with $\s ^0$ being the identity. 

Then the unbroken generators of $U(6)$ are: $\l ^0\otimes \s ^0, \l ^0 \otimes \s^3$ are the two unbroken $U(1)$'s. Schematically (the labels 1,2 stand for the two branes):
\[
U(1)_1=\left( \begin{array}{cc} \l^0 & 0\\0 & 0 \end{array} \right);~~~U(1)_2=\left( \begin{array}{cc} 0 & 0\\0 & \l^0 \end{array} \right)
\] or equivalently:
\be  \label{1}
U(1)_B=\left( \begin{array}{cc} \l^0 & 0\\0 & \l^0 \end{array} \right);~~~U(1)_{3}: 2t_3=\left( \begin{array}{cc} \l^0 & 0\\0 & -\l^0 \end{array} \right)
\ee
Similarly
$\l ^a\otimes (\s ^0+\s ^3)/2, \l ^a \otimes (\s^0-\s^3)/2,~~a~= 1,2...8$ are the two $SU(3)$. These are not excited in our condensate solutions. They do contribute in an instanton contribution localized at $u=u_c$. They generate the Chern Simons interaction which gives a source at $u=u_c$. Schematically:
\be
\left( \begin{array}{cc} \l^a & 0\\0 & 0 \end{array} \right);~~~\left( \begin{array}{cc} 0& 0\\0 & -\l^a \end{array} \right)
\ee

The eighteen broken generators are:

$\l^0\otimes \s^{1,2}$ are the generators corresponding to the $W^{1,2}$ massive gauge fields of the $SU(2)$ that condense. 
Schematically
\be   
t_1=\left( \begin{array}{cc} 0 &\l^ 0\\\l^0 & 0 \end{array} \right);~~~t_2=\left( \begin{array}{cc} 0& -i \l^0\\i \l^0 & 0 \end{array} \right)
\ee

Finally the sixteen $\l^a \otimes \s ^{1,2}$ are the broken generators corresponding to gauge fields that are not excited. Schematically:
\be
\left( \begin{array}{cc} 0 &\l^ a\\\l^a & 0 \end{array} \right);~~~\left( \begin{array}{cc} 0& -i \l^a\\i \l^a & 0 \end{array} \right)
\ee

The total number of generators is 36 as required for $U(6)$. This is easily generalized to $U(2N_d)$. 

The Yang-Mills analysis in this paper involves $t_1,t_2,t_3$. As can be seen they see nothing of the $SU(3)$ internal structure. Hence these indices are suppressed in all the equations of this paper. 

Thus for each set of three coincident D8 branes the DBI analysis given in this paper follows that of \cite{SS1,BLL}.

\section{Symmetrized Trace Prescription}\label{appbs1}


In this appendix we give a useful result on symmetrized trace
(Str) prescription.  This is useful in the two flavor case when
there is only the $0$ component of the $U(2)$ Yang-Mills fields
being non-zero and the branes are on top of each other (so that
the symmetric ``metric" part of the Dirac-Born-Infeld is proportional
to the identity).  Evaluate using symmetrized trace prescription
the object
\[
e^{ {1 \over 2} \; Tr \; ln (1 + F_3 + F) } \; \; . 
\] 
The prescription is to expand in a power series in $F_3+F$ and
symmetrize each term.

Let $F=F^+ \tau ^+ + F^- \tau ^-$ and 
$F_3 = {1 \over 2} \; F^3 \tau ^3 \;$.

Observations:

\begin{enumerate}

\item 
Since $F_{u0}=-F_{0u}$ is the only non zero component, the trace
(over Lorentz indices) of an odd number of $F$'s is zero.

\item 
\[ 
F_3 ^2 = {(F^3)^2 \over 4} I
\] 
\[ 
F^2 = F^+ F^- \{\tau ^+ ,\tau ^- \}= F^+F^-I
\] 
\[
F_3 F + F F_3 = 0 \; \; . 
\]

\item 
Consider a term $F_3^n F^m$. It has to be symmetrized completely
so we have all permutations.  Consider the first two places: If
it is $F_3 F .....$ then one has to add the permutation $F F_3
....$ where the three dots are identical in both cases. This
will give $F_3 F...+FF_3... = \{ F_3 , F\} ... = 0 \;$. Thus we
can conclude that in the first position either there should be
two $F$'s or two $F_3$'s. Having done this operation on the
first two places, we consider the next two places. The same
argument holds. This can be repeated.

Thus we conclude that $Str ~(F_3^n F^m) \approx (F^2)^{m \over
2} (F_3 ^2)^{n \over 2} \;$. (Here $F_3^2 = {(F^3)^2 \over 4}$
and $F^2 = F^+ F^-$, {\em i.e.}  without any matrices.) We need
to determine the precise coefficient. There are $ (n+m)! \over
n!  m!$ permutations. So
\[
Str ~ (F_3^n F^m)  = {n! m! \over (n+m)!} \; \; 
Tr \sum_{Perm} (F_3^n F^m) \; \; . 
\]
The number of non zero terms is the number of ways we can pick
${m \over 2}$ places to place the $F^2$ in a total of $m + n
\over 2$ places (i.e of the form $F^2 F_3^2 F^2 F^2
F_3^2.... \;$). This is ${m + n \over 2}! \over {n \over 2}!{m
\over 2}! \;$. So the final answer is (A factor of 2 for the
trace)
\[
Str (F_3^n F^m) = 2 
{ {m + n \over 2}! \over {n \over 2}! {m \over 2}! } \; 
{n! m! \over (n + m)!} \; 
(F^2)^{m \over 2} (F_3^2)^{n \over 2} \; \; . 
\]

\item 
Based on the above $n,m$ have to be even.

\end{enumerate}

Our starting point is 
\be 
e^{{1 \over 2} \; Tr \; ln (1 + F_3 + F)}
\ee 
\[ 
= Exp \left[ {1 \over 2} \left( -2 ( 
{(F_3 + F)^2 \over 2} + { ((F_3 + F)^2)^2 \over 4}
+ { ((F_3+F)^2)^3 \over 6} 
+... ) \right) \right]
\] 
using the fact that odd powers vanish. This is the power series
expansion of $\sqrt{1- (F + F_3)^2} \;$. So let us define $C_n$
by:
\[
\sqrt{1 - y^2} = \sum_{n = 0}^\infty C_n (y^2)^n \; \; .
\]
Then
\[
e^{ {1 \over 2} \; Tr \; ln (1 + F_3 +F)} 
= \sum_n C_n ((F + F_3)^2)^n \; \; . 
\] 
Thus the coefficient of a term $F_3^{n} F^m$ in the above is
$C_{n + m \over 2} \; {(n + m)! \over n! m!} \;$. Thus using the
formula for symmetrized trace derived above we get:
\[
C_{n + m \over 2} \; {(n + m)! \over n! m!} \; \; 
2 {{m + n \over 2}! \over {n \over 2}! {m \over 2}!}
\; \; {n! m! \over (n + m)!} \; 
(F^2)^{m \over 2} (F_3^2)^{n \over 2}
\]
\[
= C_{n + m \over 2} \; \; 
2 {{m + n \over 2}! \over {n \over 2}! {m \over 2}!} \; 
(F^2)^{m \over 2} (F_3^2)^{n \over 2} \; \; . 
\]

We can now use it to get coefficient of powers of $F^2 \;$:

{\bf Case 1 $\; m = 0 \;$}:
This gives 
\[ 
2 \; C_{n \over 2} \; (F_3^2)^{n \over 2} \; \; . 
\] 
Summing over $n$ (even) gives $2 \sqrt{1 - F_3^2} \;$. 

{\bf Case 2 $\; m = 2 \;$}:
\[ 
2 \; C_{{n \over 2} + 1} \; ({n \over 2} + 1) \; 
F^2 \; (F_3^2)^{{n \over 2}} \; \; . 
\]

We need to perform the sum (write $n=2n_1$)
\[ 
2 F^2 \sum_{n_1 = 0}^\infty 
C_{n_1+1} (n_1 + 1) (F_3^2)^{n_1} \; \; . 
\] 
Let us rescale $F_3^2 \rightarrow x F_3^2 \;$. Then we get:
\[ 
2 F^2 \sum_{n_1} C_{n_1 + 1} (n_1 + 1) x^{n_1} (F_3^2)^{n_1}
\] 
which can be rewritten as 
\[ 
{d \over d x} \; 
2 F^2 \sum_{n_1} C_{n_1 + 1} x^{n_1 + 1} (F_3^2)^{n_1}
\]
\[
= {d \over d x} \; 
2 {F^2 \over F_3^2} \sum_{n_1} C_{n_1 + 1}
x^{n_1 + 1} (F_3^2)^{n_1 + 1} 
\] 
\[
= {d \over d x} \; 
2 {F^2 \over F_3^2} \sum_{n_2 = 1}^\infty C_{n_2}
x^{n_2} (F_3^2)^{n_2} 
\] 
where, in the last expression, $n_2 = n_1 + 1 \;$. We can extend
the lower limit of sum to $n_2 = 0$ by the addition of an
$x$-independent term without any change in it's value. Thus we
get
\[  
{d \over dx} 2 {F^2 \over F_3^2} \sum_{n_2 = 0}^\infty 
C_{n_2} x^{n_2} (F_3^2)^{n_2}
\]
\[
= {d \over dx} 2 {F^2 \over F_3^2} \sqrt{1 - x F_3^2}
\]
\[
= - {F^2 \over \sqrt{1 - F_3^2}} 
\] 
(setting $x = 1 \;$). 

Thus the first two terms in the power series for $F^2$ is
\[ 
2 \sqrt{1 - F_3^2} - {F^2 \over \sqrt{1 - F_3^2}} \; \; . 
\]

{\bf General Case:}

Let us consider the general term with $(F^2)^{m\over 2}$. Let
${n\over 2} = n_1$ and ${m\over 2}=m_1$. Then we have
\[
2 C_{n_1+m_1} {(n_1+m_1)!\over n_1! m_1!}(F^2)^{m_1}(F_3^2)^{n_1}
\]
\[
= 2(F^2)^{m_1}C_{n_1+m_1} {(n_1+m_1)(n_1+m_1-1)....(n_1+1)\over
m_1!} (F_3^2)^{n_1} \; \; . 
\]
Replacing $F_3^2 \rightarrow x F_3^2$ as before we can write
this as:
\[ 
2(F^2)^{m_1}C_{n_1+m_1}{1\over m_1!}{ d^{m_1}\over dx^{m_1}}
x^{n_1+m_1}(F_3^2)^{n_1}
\]
\[
=2({F^2\over F_3^2})^{m_1}C_{n_1+m_1}{1\over m_1!}{ d^{m_1}\over
dx^{m_1}} x^{n_1+m_1}(F_3^2)^{n_1+m_1} \; \; . 
\]
Do the sum over $n_1$:
\[
=2({F^2\over F_3^2})^{m_1}{1\over m_1!}{ d^{m_1}\over dx^{m_1}}
\sum_{n_1=0}^{\infty} C_{n_1+m_1}x^{n_1+m_1}(F_3^2)^{n_1+m_1}
\; \; . 
\] 
Let $n_1+m_1=n_2$. Then we get:
\[
=2({F^2\over F_3^2})^{m_1}{1\over m_1!}{ d^{m_1}\over
dx^{m_1}}\sum _{n_2=m_1}^{\infty}C_{n_2}x^{n_2}(F_3^2)^{n_2}
\; \; . 
\]
Because of the derivative we can extend the sum over $n_2$ to 0.
\[
=2({F^2\over F_3^2})^{m_1}{1\over m_1!}{ d^{m_1}\over
dx^{m_1}}\sum _{n_2=0}^{\infty}C_{n_2}x^{n_2}(F_3^2)^{n_2}
\]
\be
=2({F^2\over F_3^2})^{m_1}{1\over m_1!}{ d^{m_1}\over
dx^{m_1}}\sqrt{1-xF_3^2} \; \; . 
\ee
Summing over $m_1$ one finds a Taylor series for 
\be
2\sqrt{1-{(F^3)^2\over 4}-F^+F^-} \; \; . 
\ee

Actually one can conclude that this had to be so from the first
term in the Taylor series, (the case $m=2$) and requiring
Lorentz invariance of the final expression, which uniquely fixes
the expression inside the square root.


\section{$N_f$ number of $D8$ branes: Action}\label{appkr2}


In this Appendix, we consider the non abelian generalization of
the action $S_{DBI}$ for $N_f$ number of $D8$ branes. We set
$\sigma = u \;$. Thus, the worldvolume coordinates $\sigma^\mu$
are
\[
\sigma^\mu = (x^0, u, x^1, x^2, x^3, \theta^1, \cdots,
\theta^4) \; \; . 
\]
The $D8$ brane action may now be written as
\[
S_{DBI} = - \tilde{C_9} \int d^9 \sigma \; g_s e^{- \Phi} \; 
Str \; \sqrt{ - det \; (G + M) }
\]
where the components $G_{\mu \nu}$ and $M_{\mu \nu}$ of the
matrices $G$ and $M$ are given by
\begin{eqnarray*}
G_{\mu \nu} & = &  g_{\mu \nu} 
+ g_{\tau \tau} \; \partial_\mu \tau \; \partial_\nu \tau \;
+ F_{\mu \nu} \\
M_{\mu \nu} & = &  g_{\tau \tau} \; 
{\cal D}_\mu \phi \; {\cal D}_\nu \phi \; + f_{\mu \nu}
\end{eqnarray*}
with $g_{M N}$ the background metric of $D4$ branes. In the
above expressions, we have split the $U(N_f)$ terms into the
abelian $U(1)$ terms $\partial_\mu \tau$ and $F_{\mu \nu} \;$,
and into the non abelian $SU(N_f)$ terms ${\cal D}_\mu \phi = (
{\cal D}_\mu \phi^a ) \; t_a \;$, $\; \phi = \phi^a \; t_a \;$,
and $f_{\mu \nu} = f_{\mu \nu}^a \; t_a $ where $t_a$ are the
$SU(N_f)$ generators satisfying the algebra $[t_b, \; t_c] = i
\; C^a_{\; \; b c} \; t_a \;$ and
\begin{eqnarray}
{\cal D}_\mu \phi^a & = & \partial_\mu \phi^a 
+ \; g \; C^a_{\; \; b c} \; A_\mu^b \; \phi^c  \nonumber \\
f_{\mu \nu}^a & = & 
\partial_\mu A_\nu^a - \partial_\nu A_\mu^a 
+ \; g \; C^a_{\; \; b c} \; A_\mu^b \; A_\nu^c \; \; . 
\label{covder}
\end{eqnarray}
The $SU(N_f)$ invariant action is to be obtained by symmetrized
trace ($Str$) prescription. In this paper, $g = - 1 \;$.

In this paper, we will keep $\tau$ and $F_{\mu \nu}$ terms
fully, but keep the non abelian $\phi$ and $f_{\mu \nu} \;$
terms only up to quadratic order. We have
\begin{eqnarray}
\frac{\sqrt{det \; (G + M)}}{\sqrt{det \; G }}
& = & 1 \; + \; \frac{1}{2} \; G^{\mu \nu} M_{\nu \mu} \; 
+ \; \frac{1}{8} \; ( G^{\mu \nu} \; M_{\nu \mu} )^2 
\nonumber \\
& & - \; \frac{1}{4} \; G^{\mu \nu} G^{\lambda \sigma} \; 
M_{\nu \lambda} M_{\sigma \mu} \; + \; {\cal O} (M^3) 
\label{matrix}
\end{eqnarray}
where $G^{\mu \nu}$ are the components of the matrix $G^{- 1}
\;$. Let
\[
G_\pm^{\mu \nu} = \frac{1}{2} \; (G^{\mu \nu} \pm G^{\nu \mu})
\]
denote the symmetric and antisymmetric part of $G^{\mu \nu} \;$.
Taking the symmetrized trace over the group indices we then have
\begin{eqnarray*}
G^{\mu \nu} M_{\nu \mu} & = & G_+^{\mu \nu} \; 
g_{\tau \tau} \; {\cal D}_\mu \phi^a \; {\cal D}_\nu \phi^a \\
( G^{\mu \nu} \; M_{\nu \mu} )^2 & = & 
( G_-^{\mu \nu} \; f_{\mu \nu}^a )^2 \\ 
G^{\mu \nu} G^{\lambda \sigma} \; 
M_{\nu \lambda} M_{\sigma \mu} & = & 
- \; ( G_+^{\mu \nu} G_+^{\lambda \sigma} \; 
- \; G_-^{\mu \nu} G_-^{\lambda \sigma} ) \; 
f_{\mu \lambda}^a f_{\nu \sigma}^a 
\end{eqnarray*}
up to quadratic order in $\phi^a$ and $f_{\mu \nu}^a \;$. The
negative signs in the last expression arise because of the
shuffling of $\mu, \nu, \cdots$ indices. We write the resulting
action as
\begin{equation}\label{sbi}
S_{DBI} = - \tilde{C_9} \int d^9 \sigma \; \gamma \; {\cal L}_*
\; \; \; , \; \; \; \; 
\gamma = g_s e^{- \Phi} \; \sqrt{- det \; G} 
\end{equation}
with ${\cal L}_* \;$ given, up to quadratic order in $\phi^a$ and
$f_{\mu \nu}^a \;$, by
\begin{eqnarray}
{\cal L}_* & = & 1 + \frac{1}{2} \; G_+^{\mu \nu} \; 
g_{\tau \tau} \; {\cal D}_\mu \phi^a \; {\cal D}_\nu \phi^a
+ \frac{1}{8} \; ( G_-^{\mu \nu} \; f_{\mu \nu}^a )^2 
\nonumber \\
& & + \; \frac{1}{4} \; ( G_+^{\mu \nu} G_+^{\lambda \sigma} \; 
- G_-^{\mu \nu} G_-^{\lambda \sigma} ) \; 
f_{\mu \lambda}^a f_{\nu \sigma}^a \; \; . \label{l*}
\end{eqnarray}

The equations of motion for the fields $(\phi^a , \; A_\mu^a)
\;$ can now be obtained. The $G_-$ terms cancel each other if
the rank of the matrix $G_-$ is two (which is the present case)
or three. Ignoring therefore the $G_-$ terms, the equations of
motion are given by
\begin{eqnarray} 
& & 
{\cal D}_\mu \left( \gamma \; G_+^{\mu \nu} \; g_{\tau \tau} \;
{\cal D}_\nu \phi^a \right) \; = \; 0 \label{eomphia} \\
& & 
{\cal D}_\mu \left( \gamma \; G_+^{\mu \nu} G_+^{\lambda \sigma}
\; f_{\nu \sigma}^a \right) \; = \; g \; C^a_{\; \; b c} 
\; \phi^b \; 
\left( \gamma \; G_+^{\lambda \sigma} \; g_{\tau \tau} \;
{\cal D}_\sigma \phi^c \right) \label{eomfa}
\end{eqnarray}
where $ {\cal D}_\mu (*^a) = \partial_\mu (*^a) + \; g \;
C^a_{\; \; b c} \; A_\mu^b \; (*^c) \;$.

In our case, $a_0$ is the only non vanishing component of the
$U(1)$ gauge field and $(\tau, a_0)$ depend on $u$ only. Hence
$F_{0 u} = - (a_0)_u \; $,
\[
G_{u u} = g_{u u} + g_{\tau \tau} \; \tau_u^2 
\; \; \; , \; \; \;
G_{0 u} = - G_{u 0} = F_{0 u} \; \; ,
\]
and $G_{\mu \nu} = g_{\mu \nu}$ otherwise. The subscripts $u$ on
$(\tau, a_0)$ denote their \\ $u-$derivatives. $G^{\mu \nu}$ and
$G_\pm^{\mu \nu} \;$ are now given by
\begin{eqnarray}
& & G^{0 0} = G_+^{0 0} = \frac{g^{0 0}}{\Delta} \; 
(1 + g^{u u} g_{\tau \tau} \; \tau_u^2) 
\; \; \; \; , \; \; \; \; \; \; 
G^{u u} = G_+^{u u} = \frac{g^{u u}}{\Delta} \nonumber \\
& & G^{0 u} = - G^{u 0} = G_-^{0 u} \; = \; 
\frac{g^{0 0} \; g^{u u} }{\Delta} \; \; F_{0 u} 
\label{g+}
\end{eqnarray}
and $G^{\mu \nu} = G_+^{\mu \nu} = g^{\mu \nu}$ and $G_-^{\mu
\nu} = 0 \;$ otherwise. $\Delta \;$ in these equations is
defined in equation (\ref{Delta}), namely
\[
\Delta = 1 + g^{u u} g_{\tau \tau} \; \tau_u^2 
+ g^{0 0} g^{u u} \; F_{0 u}^2 \; \; .
\]
With these substitutions, and noting that 
\begin{equation}\label{gammad4}
\gamma = g_s e^{- \Phi} \; \sqrt{- det \; G} 
 = u^4 \; \sqrt{\frac{H}{f}} \; \sqrt{\Delta} 
\end{equation}
for the $D4$ brane background given in equation
(\ref{d4metric}), it can be checked that the action $S_{DBI} \;$
above gives the action in equation (\ref{49}). For the $D4$
brane background, we also have
\begin{eqnarray}
\gamma & = & 
\left( \frac{u^8 f}{D} \right) \frac{\sqrt{H D}}{f}
\nonumber \\
G_+^{0 0} & = & 
- \left( \frac{D + c_0^2}{u^8 f} \right) \sqrt{H}
\nonumber \\
G_+^{u u} & = & 
\left( \frac{D}{u^8 f} \right) \frac{f}{\sqrt{H}}
\label{g+0u}
\end{eqnarray}
where $D = (u^8 + \frac{d_0^2}{H}) f - c_0^2 \;$ and the terms
inside the brackets above $\to 1 \;$ for large $u \;$.

The analysis leading to the action given in (\ref{sbi}) is also
applicable to the case where, instead of non abelian fields, one
switches on other component(s) of the abelian $U(1)$ field and
studies their leading order fluctuations. For example, let $a_2
(t, u) \;$ be the $U(1)$ gauge field along the $\sigma^2 = x^1$
direction in the $D8$ brane world volume. Then the action is
given, up to quadratic order in $a_2 \;$, by
\begin{eqnarray}
S(a_2) & \sim & \int \gamma \; 
G_+^{\mu \nu} G_+^{\lambda \sigma} \; 
(\partial_\mu a_\lambda - \partial_\lambda a_\mu) \;  
(\partial_\nu a_\sigma - \partial_\sigma a_\nu) \;  
\nonumber \\ 
& \sim & \int \gamma \; 
(G_+^{0 0} \; G_+^{2 2} \;  (\partial_t a_2)^2 
+ G_+^{u u} \; G_+^{2 2} \; (\partial_u a_2)^2 ) \; \; . 
\end{eqnarray}
The second line follows since $a_2$ is a function of $(\sigma^0,
\sigma^1) = (t, u)$ only. Using equation (\ref{g+0u}) and
$G_+^{2 2} = g^{2 2} = \sqrt{H} \;$, we get the action 
given in (\ref{71}) 
\[
S(a_2) \sim \int \sqrt{H D} \; 
\left( - \frac{H (D + c_0^2)}{f D} \; (\partial_t a_2)^2 
+ (\partial_u a_2)^2 \right) 
\]
since $\frac{H (D + c_0^2)}{f} = (u^5 + d_0^2) \;$.



\section{$N_f \; (= 2)$ number of $D8$ branes: \\ 
Equations of motion}\label{appkr3}


In this Appendix, we write down the general equations of motion
which follow from the action given in equation (\ref{sbi}).
These equations may then be specialized to the various cases
studied in the paper.

Consider the case where $A_\mu^a = 0$ if $\mu \ne 0 , 1 \;$ and
the fields $(A_0^a, A_1^a, \phi^a) \;$ depend only on
$(\sigma^0, \sigma^1) = (t, \; u) \;$. Then $f_{\mu \nu}^a = 0
\;$ if $\{\mu, \nu\} \ne \{0, 1\} \;$. Taking $G_+^{\mu \nu} \;$
to be diagonal and defining \footnote{
For the $D4$ brane background, using $g_{\tau \tau} =
\frac{f}{\sqrt{H}} \;$ and equations (\ref{g+0u}), we have
\[
(\gamma^0, \gamma^1, \gamma^f) = \sqrt{H D} \; 
\left( - \frac{D + c_0^2}{D}, \; \frac{f}{H} \; 
- \frac{D + c_0^2}{u^8 f} \right) \; \; .
\]
Note that $\gamma^1 \gg | \gamma^0 | \sim | \gamma^f | \;$ for
large $u \;$ since $H = \frac{1}{u^3} \;$.} 
\[
\gamma^0 = \gamma \; G_+^{0 0} \; g_{\tau \tau} 
\; \; , \; \; \; 
\gamma^1 = \gamma \; G_+^{u u} \; g_{\tau \tau} 
\; \; , \; \; \; 
\gamma^f = \gamma \; G_+^{0 0} G_+^{u u}
\]
the equations of motion (\ref{eomphia}) and (\ref{eomfa}) for
the fields $(\phi^a, A_0^a, A_1^a) \;$ become
\begin{eqnarray}
{\cal D}_0 (\gamma^0 \; {\cal D}_0 \phi^a) \; 
+ \; {\cal D}_1 (\gamma^1 \; {\cal D}_1 \phi^a) & = & 0 \\
{\cal D}_0 (\gamma^f \; f_{0 1}^a) \; - \; g \; C^a_{\; \; b c}
\; \phi^b \; (\gamma^1 \; {\cal D}_1 \phi^c) & = & 0 \\
{\cal D}_1 (\gamma^f \; f_{0 1}^a) \; + \; g \; C^a_{\; \; b c}
\; \phi^b \; (\gamma^0 \; {\cal D}_0 \phi^c) & = & 0 \; \; .
\end{eqnarray}

We now specialize to the case of $SU(2) \;$, namely $N_f = 2
\;$. We choose the $SU(2)$ generators to be $(t_+, \; t_-, \;
t_3) \;$ with the independent non vanishing components of the
structure constants $C^a_{\; \; b c} = - \; C^a_{\; \; c b} \;$
given by
\[
C^+_{\; \; 3 +} = - \; i 
\; \; , \; \; \; 
C^-_{\; \; 3 -} = + \; i 
\; \; , \; \; \; 
C^3_{\; \; + -} = - \; 2 i \; \; . 
\]
The fields $(\phi^a, A^a_\mu) \;$ are given by 
\begin{eqnarray*}
\phi = \phi^a \; t_a & = & 
\phi^+ \; t_+ + \phi^- \; t_- + \phi^3 \; t_3 \\
A_\mu = A_\mu^a \; t_a & = & 
A_\mu^+ \; t_+ + A_\mu^- \; t_- + A_\mu^3 \; t_3 
\end{eqnarray*}
where $(\phi^+, A^+_\mu) \;$ are complex fields, $(\phi^-,
A^-_\mu) \;$ their complex conjugates, and $(\phi^3, A^3_\mu)
\;$ are real fields. Then ${\cal D}_\mu \phi^a \;$ are given by
\begin{eqnarray}
{\cal D}_\mu \phi^\pm & = & \partial_\mu \phi^\pm 
\; \mp \; i g \; (A_\mu^3 \phi^\pm - A_\mu^\pm \phi^3) \\
{\cal D}_\mu \phi^3 & = & \partial_\mu \phi^3
\; -  \; 2 i g \; (A_\mu^+ \phi^- - A_\mu^- \phi^+) 
\end{eqnarray}
and $f_{\mu \nu}^a \;$ by 
\begin{eqnarray}
f_{\mu \nu}^\pm & = & 
\partial_\mu A_\nu^\pm - \partial_\nu A_\mu^\pm \; 
\mp \; i g \; (A_\mu^3 A_\nu^\pm - A_\mu^\pm A_\nu^3) \\
f_{\mu \nu}^3 & = & 
\partial_\mu A_\nu^3 - \partial_\nu A_\mu^3 \; 
- \; 2 i g \; (A_\mu^+ A_\nu^- - A_\mu^- A_\nu^+) \; \; . 
\end{eqnarray}
We further choose a gauge where $A_0^\pm = A_1^3 = 0 \;$.
($A^a_1 \;$ corresponds to the gauge field along $\sigma^1 = u
\;$ direction.) The non vanishing gauge field components are
then $(A_0^3, A_1^\pm) \equiv (A, w^\pm) \;$. Then
\begin{eqnarray}
({\cal D}_0 \phi^\pm \; , \; \; {\cal D}_0 \phi^3 ) & = &
(\phi^\pm_t \mp i g A \phi^\pm \; , \; \; \phi^3_t ) \\
({\cal D}_1 \phi^\pm \; , \; \; {\cal D}_1 \phi^3) & = & 
(\phi^\pm_u \pm i g w^\pm \phi^3 \; , \; \; 
\phi^3_u - 2 i g (w^+ \phi^- - w^- \phi^+) ) \nonumber \\
\\
(f_{0 1}^\pm \; , \; \; f_{0 1}^3) & = & 
(w^\pm_t \mp i g A w^\pm \; , \; \;  - A_u)
\end{eqnarray}
where the subscripts $t, u$ on the fields $(\phi^a, A, w^\pm)$
denote their derivatives with respect to $t, u \;$. 

The equations of motion (\ref{eomphia}) and (\ref{eomfa}) for $a
= 3$ become
\begin{eqnarray}
& & (\gamma^0 \phi^3_t)_t + (\gamma^1 \phi^3_u)_u 
- 2 i g (\gamma^1 (w^+ \phi^- - w^- \phi^+))_u 
\nonumber \\
& & - 2 i g \gamma^1 (w^+ \phi^-_u - w^- \phi^+_u) 
- 4 g^2 \gamma^1 w^+ w^- \phi^3 \; = \; 0 
\label{3} \\ \nonumber \\ 
& & - (\gamma^f A_u)_t 
+ 2 i g \gamma^1 (\phi^+ \phi^-_u - \phi^- \phi^+_u) 
\nonumber \\
& & + 2 g^2 \gamma^1 (w^+ \phi^- + w^- \phi^+) \phi^3
\; = \; 0 \label{13} \\ \nonumber \\
& & - (\gamma^f A_u)_u - 2 i g \gamma^f (w^+ w^-_t - w^- w^+_t)
- 2 i g \gamma^0 (\phi^+ \phi^-_t - \phi^- \phi^+_t) 
\nonumber \\
& & + 4 g^2 (\gamma^f w^+ w^- + \gamma^0 \phi^+ \phi^-) A
\; = \; 0 \; \; , \label{03} 
\end{eqnarray}
and, for $a = + \;$, they become
\begin{eqnarray}
& & (\gamma^0 (\phi^+_t - i g A \phi^+))_t 
+ (\gamma^1 (\phi^+_u + i g w^+ \phi^3))_u 
- i g \gamma^0 A \phi^+_t 
\nonumber \\
& & + i g \gamma^1 w^+ \phi^3_u 
+ 2 g^2 \gamma^1 (w^+ \phi^- - w^- \phi^+) w^+ 
- g^2 \gamma^0 A^2 \phi^+ \; = \; 0 
\label{+} \nonumber \\ \\
& & (\gamma^f (w^+_t - i g A w^+))_t - i g \gamma^f A w^+_t
+ i g \gamma^1 (\phi^3 \phi^+_u - \phi^+ \phi^3_u)
\nonumber \\ 
& & - 2 g^2 \gamma^1 (w^+ \phi^- - w^- \phi^+) \phi^+ 
- g^2 (\gamma^1 (\phi^3)^2 + \gamma^f A^2) w^+ \; = \; 0 
\label{1+} \nonumber \\ \\ 
& & (\gamma^f w^+_t)_u 
- i g \gamma^0 (\phi^3 \phi^+_t - \phi^+ \phi^3_t) 
- i g (\gamma^f w^+ A)_u 
\nonumber \\
& & - i g \gamma^f w^+ A_u - g^2 \gamma^0 A \phi^3 \phi^+ 
\; = \; 0 \; \; , \label{0+}
\end{eqnarray}
and the $a = -$ equations of motion are the complex conjugates
of the $a = +$ ones. In the static case, all $t-$derivatives
vanish. Let $\phi^\pm = \phi \; e^{\pm i \theta} \;$ and $w^\pm
= w \; e^{\pm i \Omega} \;$ where $\phi$ and $w$ are real. It
then follows from equations (\ref{13}) and (\ref{1+}) that
$\Omega = \theta + \frac{\pi}{2} = constant \;$.

In this paper, $g = - 1 \;$. Setting $\gamma^0 = \gamma^f = - 1
\;$ and $\gamma^1 = 1 \;$ and charged scalars to zero, gives the flat space equations of
motion (\ref{fs1}) -- (\ref{fs5}). With $\gamma^0 = \sqrt{- g}
g^{0 0} g_{\tau \tau} \;$, $\gamma^1 = \sqrt{- g} g^{u u}
g_{\tau \tau} \;$, and $\gamma^f = \sqrt{- g} g^{0 0} g^{u u}
\;$, and with all $t-$derivatives vanishing, the above equations
give static curved space equations of motions (\ref{50}).

\section{Flat Space Equations}
Two solutions were given in the text. The equations of motion
are given here. The field $\chi$ was defined in Section 3. If below we assume that $\chi =\phi_1$ is real then we have the
solution with one adjoint scalar corresponding to the D8 brane case. If we leave $\chi$
complex the equations could be used for other branes.

The equations of motion, in the same gauge as above, and with
time derivatives set to zero, are:
\be
{\mathbf {\delta S \over \delta A_0^3}} = 
- \du^2 A_0^3 + 2 \; (2 w^+ w^- + \bar \chi^+ \chi^- 
+ \bar \chi^- \chi^+) \; A_0^3
\ee
\[
{\mathbf {\delta S \over \delta A_u^3}} = 
\chi^+ \du \bar \chi^- - \chi^- \du \bar \chi^+ 
+ \bar \chi^+ \du \chi^- - \bar \chi^- \du \chi^+
\]
\be
+ i w^+ \; (\chi^3 \bar \chi^- + \bar \chi^3 \chi^-)
+ i w^- \; (\chi^3 \bar \chi^+ + \bar \chi ^3 \chi ^+)
\ee
\[
{\mathbf {\delta S \over \delta \chi^3}} = 
\hf \du^2 \bar \chi^3 
+ 2 i \; (w^+ \du \bar \chi^- - w^- \du\bar \chi^+)
+ i \; (\bar \chi^- \du w^+ - \bar \chi^+ \du w^-) 
\]
\be
-\hf \; \left[ \bar \chi^+ (\bar\chi^3 \chi^- 
- \chi^3 \bar \chi^-)
+ \bar \chi^- (\bar \chi^3 \chi^+ - \chi^3 \bar\chi^+) \right]
- 2 w^+w^- \; \bar \chi^3 
\ee
\be
{\mathbf {\delta S \over \delta A_0^-}} = 
2 i \; \du (w^+ A_0^3) + 2 i \; w^+ \du A_0^3 
- (\chi^3 \bar \chi^+ + \bar \chi^3 \chi^+) \; A_0^3
\ee
\[
{\mathbf {\delta S \over \delta A_u^-}} = 
i \; (\bar\chi^+ \du \chi^3 - \chi^3 \du \bar\chi^+) 
+ i \; (\chi^+ \du \bar\chi^3 - \bar\chi^3 \du \chi^+) 
\]
\be
+ \; 2 \; ((A_0^3)^2 - \bar \chi^+ \chi^- 
- \bar \chi^- \chi^+  - \chi^3\bar \chi^3) \; w^+ 
+ 4 \; \bar \chi^+ \chi^+ \; w^- 
\ee
\[
{\mathbf {\delta S \over \delta \chi^-}} = 
\du^2 \bar\chi^+ - 2 i \; w^+ \du \bar\chi^3 
- i \; \bar\chi^3 \du w^+
+ \; 2 \; (w^+ \bar\chi^- - w^- \bar\chi^+) \; w^+ 
\]
\be
+ \; (A_0^3)^2 \; \bar\chi^+
+ \hf \; \bar\chi^3 (\bar\chi^3 \chi^+ - \chi^3 \bar\chi^+) 
- \bar\chi^+ (\chi^+ \bar\chi^- - \chi^- \bar\chi^+)
\ee

Of course if we set $\chi = 0$ we have the same solution as
before. Furthermore, if we set one of the scalar fields, such as
$\phi_1 = 0 \;$, then the system is again the earlier one and we
only have an analytic solution where $\phi_2$ is also zero. So
we try to set a different set to zero. One can try for instance
to set $\phi_1^3 = 0 = \phi_2^\pm \;$.

The $A^-_0$ equation in fact suggests this possibility: $\chi^3
= \bar \chi^3$ and $\chi^+ + \bar \chi^+ = 0 \;$. Thus the term
$(\chi^3 \bar \chi ^+ + \bar \chi^3 \chi^+)A_0^3 $ vanishes and
we get the same relation as earlier between $A_0^3$ and $w^+
\;$. Note that $\bar \chi^+ = (\chi^-)^* \;$. Thus $\chi^+ = -
(\chi ^-)^* \;$.

In the $A_u^3$ equation, the coefficient of $w$ vanishes and the
derivative terms become: $2 \chi^- \du \chi^+ - 2 \chi^+ \du
\chi^- \;$. This vanishes if the phase of $\chi^+$ is constant,
independent of $u \;$.

The $A^-_u$ equation reduces to $(A_0^3)^2 = (\chi^3)^2$ if we
choose $\chi^+ w^- = \chi^- w^+ \;$, which means $\chi^+ w^-$
should be imaginary.  Thus we let $\chi^+ = i \chi = - \chi^-$
and $w^+ = w^- = w \;$, so the phase of $w^+$ is also
$u$-independent.

Using $\chi^+ w^- = \chi^- w^+$ and the fact that the phases of
$\chi$ and $w$ are constants, we see that in the $\chi^3$
equation the derivative terms cancel. And using the reality
properties of $\chi \;$, we get the same equation as earlier:
\[ 
\hf \du^2 \chi^3 + 2 \chi^+ \chi^- \chi^3 
- 2 w^+ w^- \chi^3 = 0 \; \; . 
\] 
Note that $\chi^+\chi^-$ is negative definite.

Finally the $\chi^-$ equation reduces to $\du^2 \chi^+=0$ and $2
i w^+ \du \chi^3 + i \chi^3 \du w^+ = 0 \;$. We impose these two
separately, because $\chi^3 $ has to satisfy the same equation
as $A_0^3 \;$.

Thus, when all the dust settles we have a system of equations,
very similar to the earlier one, except that $\chi^+$ can be
nonzero: either a constant or linear in $u$. But this changes
fairly dramatically the behavior of $A_0^3$ - it becomes
exponential rather than linear. If we take $\chi^+ = \chi_0 =
constant$ then $A_0^3$ can be solved for in closed form: One
finds using the same methods as earlier
\be
A_0^3(u) = \sqrt{a \chi_0} \; cosh ~(\chi_0 \; u)
\ee 
Note that $\du A^3_0 (0) = 0 \;$, and thus the electric field
vanishes at $u = 0$ as required by symmetry.


\section{$N_f \; (= 2)$ number of $D8$ branes: \\ 
Set up for the study of Stability}\label{appkr4}


In this Appendix, we write down the general equations of motion
for small field fluctuations in the background of non zero
$A^3_0 \;$ and $\phi^3 \;$. They may then be specialized to the
case studied in the paper.

Consider solutions to the equations (\ref{3}) -- (\ref{0+}) for
the static case and with $\phi^\pm = w^\pm = 0 \;$. Setting
$\phi^3 = B \;$, equations (\ref{3}) and (\ref{03}) give
\begin{equation}\label{qE}
(\gamma^1 B_u)_u \; = \; (\gamma^f A_u)_u \; = \; 0 \; \;. 
\end{equation}
Hence $\gamma^1 B_u = q$ and $\gamma^f A_u = E$ where $q$ and
$E$ are constants. Other equations are satisfied identically.
Setting $B(u_c) = A(u_c) = 0 \;$, we have
\[
A(u)  = E \; \int^u_{u_c} \; \frac{d u}{\gamma^f (u)}  
\; \; \; \; , \; \; \; \; \; \; 
B(u) = q \; \int^u_{u_c} \; \frac{d u}{\gamma^1 (u)} \; \; .
\]

Consider the fluctuations of the fields around this static
background. Thus, we write
\[
A(t, u) = A(u) + a(t, u) 
\; \; \; , \; \; \; \; 
B(t, u) = B(u) + b(t, u) 
\]
where $A(u)$ and $B(u)$ are the static background solutions
given above. The fields $(a, w^\pm, b, \phi^\pm)$ are functions
of $(t, u) \;$ and are assumed to be small. Their equations of
motion follow from equations (\ref{3}) -- (\ref{0+}). Noting
that $(\gamma^0, \gamma^1, \gamma^f)$ depend only on $u$ and
that $A(u)$ and $B(u)$ are static background solutions, we have,
to the linear order in $(a, w^\pm, b, \phi^\pm)$,
\begin{eqnarray}
& & (\gamma^0 b_t)_t + (\gamma^1 b_u)_u 
\; = \; (\gamma^f a_u)_t \; = \; (\gamma^f a_u)_u 
\; = \; 0 \\ \nonumber \\
& & \gamma^0 \phi^+_{t t} - 2 i g \gamma^0 A \phi^+_t 
+ (\gamma^1 \phi^+_u)_u + i g (\gamma^1 B w^+)_u 
\nonumber \\
& & + i g \gamma^1 B_u w^+ - g^2 \gamma^0 A^2 \phi^+
\; = \; 0 \\ \nonumber \\
& & \gamma^f w^+_{t t} - 2 i g \gamma^f A w^+_t 
+ i g \gamma^1 (B \phi^+_u - B_u \phi^+) 
\nonumber \\
& & - g^2 (\gamma^1 B^2 + \gamma^f A^2) w^+ 
\; = \; 0 \\ \nonumber \\
& & (\gamma^f w^+_t)_u - i g \gamma^0 B \phi^+_t 
- i g (\gamma^f A w^+)_u \nonumber \\
& & - i g \gamma^f A_u w^+ - g^2 \gamma^0 A B \phi^+ 
\; = \; 0 \; \; .
\end{eqnarray}
In the above equations, $A$ and $B$ are the static background
solutions and the $t, \; u$ subscripts denote $t$ and $u$
derivatives.

The independent set of fluctuations are $(a) \;$, $(b) \;$, and
$(w^\pm, \phi^\pm) \;$. We have that $(\gamma^f a_u)$ must be
constant and, hence, it simply shifts the background constant $E
\;$. The $b$ fields are traveling wave type fluctuations. In
the following we set $a = b = 0 \;$ and take $(w^\pm, \phi^\pm)
\;$ to be given by
\[
\phi^+ (t, u) = \phi(u) e^{- i m t} 
\; \; \; , \; \; \; \; 
w^+ (t, u) = i w(u) \; e^{- i m t} 
\]
where $m$ is a constant and $(w, \phi)$ are functions of $u$
only. After a little algebra, their linearized equations of
motion given above may be written as \footnote{It can be checked
that equation (\ref{uuphi}) follows upon using equations
(\ref{qE}), (\ref{uphi}), and (\ref{uw}).}
\begin{eqnarray}
(\gamma^1 \phi_u)_u - g \; 
((\gamma^1 w)_u B + 2 \gamma^1 w B_u ) 
- \gamma^0 (m + g A)^2 \; \phi \; & = &  \; 0 
\label{uuphi} \\ \nonumber \\
g \gamma^1 \; ( B \phi_u - B_u \phi )
- ( \gamma^f (m + g A)^2 + g^2 \gamma^1 B^2 ) \; w \; & = & \; 0
\label{uphi} \\ \nonumber \\
( (\gamma^f w)_u (m + g A) + 2 g \gamma^f w A_\mu )
- g \gamma^0 (m + g A) B \; \phi \; & = & \; 0 
\; \; .  \label{uw}
\end{eqnarray}
An equation for $\phi$ alone can now be obtained. For this
purpose, let
\[
h_A = \gamma^f (m + g A)^2 \; \; , \; \; \; 
h_B = g^2 \gamma^1 B^2  \; \; , \; \; \; 
h = h_A + h_B \; \; . 
\]
Then equation (\ref{uphi}) and equation (\ref{uw}), multiplied
by $(m + g A) \;$, may be written as
\[
g \gamma^1 \; (B \phi_u - B_u \phi) - h w 
\; = \; 0 \; = \;
(w \; h_A)_u - \frac{g \gamma^0}{\gamma^f} \; h_A B \phi 
\; \; .
\]
Using the above equations or, equivalently, equations
(\ref{uuphi}) and (\ref{uphi}), it can be shown after a
straightforward algebra that
\begin{equation}\label{phiuu}
\phi_{u u} + \left( ln \; \frac{\gamma^1 h_A}{h} \right)_u \;
\phi_u - \left[ \frac{\gamma^0 h}{\gamma^f \gamma^1} 
+ \frac{B_u}{B} \; \left( ln \; \frac{h_A}{h} \right)_u \right]
\; \phi \; = \; 0 \; \; .
\end{equation}

Consider the case where $g = - 1 \;$, $\gamma^0 = \gamma^f = - 1
\;$, and $\gamma^1 = 1 \;$. Then $B = q u \;$, $A = - E u \;$,
and 
\[
h_A = - (m + E u)^2 \; \; , \; \; \; 
h_B = q^2 u^2 \; \; , \; \; \; 
h = q^2 u^2 - (m + E u)^2 \; \; . 
\]
Equations (\ref{uuphi}), (\ref{uphi}), and (\ref{uw}) become
\begin{eqnarray}
& & \phi_{u u} + \; ( w_u B + 2 w B_u) + (m - A)^2 \; \phi 
\; = \; 0 \label{balauuphi} \\ \nonumber \\
& & ( B \phi_u - B_u \phi ) 
+ \left( B^2 - (m - A)^2 \right) w \; = \; 0 \label{balauphi} 
\\ \nonumber \\
& & ( - w_u (m - A) + 2 w A_u) - (m - A) B \; \phi 
\; = \; 0  \label{balauw}
\end{eqnarray}
which are same as equations (\ref{du2phi}) -- (\ref{duw}) (Amongst these three equations only two are independent).
Equation (\ref{phiuu}) becomes
\[
\phi_{u u} - \; \frac{2 m q^2 u}{(m + E u) \; h} \; \phi_u 
- \left( h - \frac{2 m q^2}{(m + E u) \; h} \right) \; \phi 
\; = \; 0 
\]
since, now, $\gamma^1 = 1 \;$ and 
\[
\left( ln \; \frac{h_A}{h} \right)_u \; = \; \frac{h_B}{h} \;
\left( ln \; \frac{h_A}{h_B} \right)_u \; = \; 
- \; \frac{2 m q^2 u}{(m + E u) \; h} \; \; . 
\]
This is same as equation (\ref{44}) for $\phi_{u u}$ with
obvious identifications.

\end{document}